\documentclass{aa}  

\usepackage{graphicx}
\usepackage{txfonts}
\usepackage{hyperref}
\usepackage{tikz}
\usetikzlibrary{arrows.meta}

\begin{document}

   \title{Small solar system objects on highly inclined orbits}
   \subtitle{Surface colours and
   lifetimes
   }

   \author{T.~Hromakina\inst{1}
        \and
            I.~Belskaya\inst{1}
        \and
            Yu.~Krugly\inst{1}
        \and
            V.~Rumyantsev\inst{2}
        \and
            O.~Golubov\inst{1}
        \and
            I.~Kyrylenko\inst{1}
        \and
           O.~Ivanova\inst{3,4,5}
        \and
            S.~Velichko\inst{1}
         \and
            I.~Izvekova\inst{6}
        \and   
            A.~Sergeyev\inst{7, 1}
        \and    
            I.~Slyusarev\inst{1}
        \and
            I.~Molotov\inst{8}
   }
          
   \institute{V.~N.~Karazin Kharkiv National University, 4 Svobody Sq., Kharkiv, 61022, Ukraine\\
   \email{hromakina@astron.kharkov.ua}
   \and
   Crimean Astrophysical Observatory, Nauchny, Crimea
   \and
   Astronomical Institute of the Slovak Academy of Sciences, SK-05960 Tatranská Lomnica,
   Slovak Republic
    \and
    Main Astronomical Observatory of the National Academy of Sciences of Ukraine,
    27 Zabolotnoho Str., 03143 Kyiv, Ukraine
    \and
    Taras Shevchenko National University of Kyiv, Astronomical Observatory, Ukraine
    \and
    ICAMER Observatory of NASU, 27 Zabolotnoho Str., Kyiv, 03143, Ukraine
    \and
    Université Côte d'Azur, Observatoire de la Côte d'Azur, CNRS, Laboratoire Lagrange, France
    \and        
    Keldysh Institute of Applied Mathematics, RAS, Miusskaya Sq. 4, Moscow 125047, Russia
   }

   \date{---; ---}

 
  \abstract
   {Less than one percent of the discovered small solar system objects have
    highly inclined orbits ($i>60^{\circ}$), and revolve around the Sun on
   near-polar or retrograde orbits.
   The origin and evolutionary history of these objects are  not yet clear.}
   {In this work we  study the surface properties and orbital dynamics
   of selected high-inclination objects.}
   {BVRI photometric observations were performed in 2019-2020 using
   the 2.0m telescope at the Terskol Observatory and the 2.6m telescope
   at the Crimean Astrophysical Observatory.
   Additionally, we searched for high-inclination objects in the Sloan Digital Sky Survey and Pan-STARRS. 
   The dynamics of the selected objects was studied using  numerical simulations.}
   {We obtained new photometric observations of six high-inclination objects
   (468861) 2013~LU28, (517717) 2015 KZ120, 2020~EP, A/2019~U5 (A/PanSTARRS), C/2018~DO4 (Lemmon), and C/2019~O3 (Palomar).
   All of the objects have   similar $B-V$, $V-R$, $R-I$ colours, which are close to those of moderately red TNOs and grey Centaurs. 
   The photometric data that were extracted from the all-sky surveys also correspond to moderately red surfaces of high-inclination objects. 
   No signs of ultra-red material on the surface of high-inclination asteroids were found, which supports the results of previous works.
   The comet C/2018 DO4 (Lemmon) revealed some complex morphology with structures that could be associated with particles that were ejected from the cometary nucleus. Its value of the parameter $Af\rho$ is around 100~cm for the aperture size of 6~000~km.
   The value of $Af\rho$ for the hyperbolic comet C/2019~O3 (Palomar) is much larger, and is in the range from 2~000 to 3~700~cm for the aperture sizes from 25~000~km to 60~000~km.
   For objects 2013~LU28, 2015~KZ120, and 2020~EP we estimated future and past lifetimes on their orbits. 
   It appears that the orbits of considered objects are strongly chaotic, and with the available accuracy of the orbital elements no reliable predictions can be made about their distant past or future.
   The lifetimes of high-inclination objects turned out to be highly non-sensitive to the precision of the orbital elements, and to the Yarkovsky orbital drift.
   }
   {}

   \keywords{Minor planets, asteroids: general --
            Comets: general
            Kuiper belt: general --
            Oort Cloud --
            techniques: photometric --
            Surveys
               }
               
   \maketitle
%

\section{Introduction}

The vast majority of   solar system objects have prograde orbits with low inclinations. According to the Minor Planet
Center\footnote{https://minorplanetcenter.net} (MPC), 
only about 250 asteroids on near-polar ($i$ around 90$^{\circ}$)
or retrograde ($i$>90$^{\circ}$) orbits have been discovered to date.
The majority of high-inclination objects belong to Centaurs, trans-Neptunian
objects (TNOs), and distant objects on cometary orbits.
The remaining $\sim$10 \%\ are near-Earth objects (NEOs). 
To date no retrograde main belt asteroids have been discovered.
On the other hand, about a half of long-period comets (including those on parabolic and hyperbolic orbits)
that may have their origin in the Oort cloud have retrograde orbits.
Based on their inclination distribution, these comets are also designated as nearly isotropic comets \citep{Levison1996}.

\citet{Jewitt05} introduced a dynamical class  called the damocloids,
point-source objects on cometary orbits that do not show signs 
of cometary activity and have a Tisserand parameter $T_J<2$.
Based on this definition, most of the high-inclination objects can be classified as damocloids. 

The origin of very high-inclination objects is not yet clear and remains a matter of debate. 
Different explanations have been proposed for the presence of such objects in the present-day solar system.
A number of studies have been performed  to investigate the origin
and evolution of such objects by  means of the numerical integration of their orbits.
According to \citet{Gladman09}, such extreme TNOs can be outliers 
from the already explored reservoirs like the Kuiper belt and/or the outer Oort cloud. 
Conversely, they could also come from a currently unobserved reservoir, 
which can also supply the Halley-type comets \citep{Gladman09}.
\citet{Brasser2012} and \citet{Volk13} showed that
the Kuiper belt is a very unlikely source of retrograde centaurs. 
\citet{Chen2016} suggested the existence of a mostly undiscovered
population of extreme objects ($q>10$~AU, $a<100$~AU, and $i>60^{\circ}$) 
that occupy a common plane and have the same origin.
Another possible explanation is the existence of an extended scattered
disk created by a hypothetical yet undiscovered big planet \citep{Batygin16}. 
The recent paper by \citet{Namouni2020} suggests that 
high-inclination Centaurs were captured from the interstellar medium. 
This conclusion is based on their finding, to a high level of  probability,  for 
high-inclination Centaurs to be on near-polar orbits 4.5~Gyr in the past.
However, this result  was rebutted by \citet{Morbidelli2020}.

Retrograde near-Earth asteroids, according to model by
\citet{Greenstreet2012}, are produced by the main belt, with 
the majority of objects originating from 3:1 resonance with Jupiter.
Another evolutionary pathway that produces retrograde NEOs was proposed by \citet{Marcos2015}. 
They suggest that it is the 9:4 mean motion resonance with 
Jupiter that triggers the orbital flip and, as a result, production of retrograde objects. 

The results of numerical integration show that most retrograde
objects are unstable and have  short life spans of less than <1 Myr
\citep{Marcos14, Li2019,Kankiewicz2016}. \citet{Kankiewicz2016}
also pointed out that the influence of the Yarkovsky effect can
potentially be important, especially on Myr timescales.

Mean motion resonances with planets play an important role in the evolution
of particular objects and in the dynamical configuration of a planetary
system as a whole \citep{Morbidelli2002, Ferraz-Mello2006, Beauge2003}.
Interestingly, \cite{Namouni2015} showed that retrograde objects are
more easily captured in mean motion resonances than prograde ones.
Various authors looked for resonances among high-inclination and retrograde objects. 
\citet{Wiegert2017} identified that object 2015~BZ509 is the first retrograde co-orbital asteroid of Jupiter. 
\citet{Connors2018} found that asteroid 2007~VW266 is in 13/14 retrograde mean motion resonance.
A retrograde TNO 2011 KT19 was found to be in 7/9 resonance with Neptune by \citet{Morais2017}.
Finally, a recent survey of resonances among retrograde asteroids was performed by \citet{Li2019}.
They identified 15 retrograde objects that are currently captured in resonances
with giant planets, and 30 retrograde objects that will be captured in resonances in the future.

Given the rarity and faintness of high-inclination objects, only for a
few of them were some physical properties  determined.
\citet{Jewitt05,Jewitt15} found that damocloids show a lack of extremely red surfaces.
\citet{Licandro18} obtained spectral data for bodies on cometary orbits.
They observed several objects on retrograde and near-polar orbits and
classified them as D type based on their visible spectra. 
\citet{Pinilla-Alonso13} obtained spectra and photometric data for two retrograde objects 2008 YB3 and 2005 VD. 
They concluded that both objects have moderately red surfaces.
The authors also investigated the dynamical evolution of these two
bodies by integrating their orbits for $10^{8}$ years into the past. 
They found that the dynamical evolution of 2005 VD is dominated by a Kozai resonance with
Jupiter \citep{Kozai1962}, while that of 2008 YB3 is dominated by close encounters with Jupiter and Saturn. 
They concluded that 2005~VD comes from the Oort cloud, whereas 2008~YB3 originated in the trans-Neptunian region. 

Here we present the first results of our observational programme dedicated to the investigation of high-inclination small solar system objects.
This work contains new photometric observations of six objects.
We also made an attempt to enlarge statistics on the visible colours of high-inclination bodies by using available data from multi-band sky surveys.
The obtained data on visible colours were compared with the published data on high-inclination objects and other dynamical classes. 
Finally, we simulated orbits of observed objects to estimate their lifetimes and to investigate the possible influence of the Yarkovsky effect.


\section{Photometric observations and data reduction}

Photometric observations were carried out at the 2.6m Shain Telescope at the Crimean Astrophysical Observatory (CrAO) and at the 2.0m Zeiss-2000 telescope at the Peak Terskol observatory in 2019-2020.
Shain telescope is equipped with a 2048$\times$2048 FLI PL-4240 CCD camera. 
The field of view is 14.4$\times$14.4 arcmin and the pixel scale is 0.84"/pxl with a 2$\times$2 binning.
The Zeiss telescope is equipped with a 2048$\times$2048 FLI PL-4301 CCD camera that covers a field of view of 10.7$\times$10.7 arcmin. 
The pixel scale is 0.62"/pxl with a 2$\times$2 binning that was applied during the observations. 
The typical seeing at CrAO ranged from 1.9 to 2.4~arcsec, and the typical seeing at the Terskol Observatory was in the range from 3.0 to 3.7~arcsec.

Image reduction was done in the standard way that included subtraction of the dark images and flat-field correction.
Aperture photometry was performed with the use of the photutils Python package \citep{Bradley2019}.
For objects that do not show signs of cometary activity, the most appropriate aperture size was determined from the full width at half maximum (FWHM) values of neighbouring stars in the image.
Several aperture sizes were used for objects with cometary activity.
Sky subtraction was performed using an annular aperture around each object's photocentre.

The absolute calibration was made using the solar-type stars in the image field.
The magnitudes of the selected stars were taken from the Pan-STARRS catalogue.
For the images that were obtained at CrAO and Terskol, all suitable stars in the image were used.
For the Pan-STARRS images, due to their large field of view, we did not use all stars in the frame. 
Instead, we only considered the stars in a 0.15~arcmin radius around the target.
On average, this radius corresponded to about 20-50 stars per image.

Our observations were performed in the standard Johnson--Cousins photometric system in BVRI broad-band filters.
The exposure times of individual images were from 90 to 180~s, depending on the brightness of the object.
Typical errors for individual frames were 0.03-0.05~mag in B filter, 0.02-0.04 in R and V filters, and 0.06-0.10~mag in I filter.

\section{Results}

Only a small number of high-inclination objects were bright enough to be observable on small ground-based telescopes during the observational runs in 2019-2020.
In total, we obtained photometric data for six objects, including three objects on retrograde orbits (2013~LU28, C/2018~DO4 (Lemmon), and A/2019~U5) and three objects on near-polar orbits (2015~KZ120, C/2019~O3 (Palomar), and 2020~EP).

Our list of potentially observable targets did not initially include active comets as one of the main aims of the programme  was to assess the surface properties of high-inclination objects. 
However, two of the observed targets appeared to be comets.
For both objects the preliminary reports on the possible cometary activity appeared during the observation planning phase and the corresponding publications in Minor Planet Electronic Circular were released soon after our observations. 
The rest of the bodies do not show any signs of cometary activity in our data.
Table~\ref{geom_table} presents the observational circumstances, such as heliocentric ($r$) and geocentric ($\Delta$) distances and phase angles ($\alpha$) together with measured and reduced magnitudes.

\begin{table*}
    \centering
    \caption{Geometrical circumstances and magnitudes of   the observed objects}
        \label{geom_table}
        \begin{tabular}{lllllllll}
        \hline
        \hline
Name&UT date & r (au)&$\Delta$ (au) &$\alpha$ (deg) & Apparent mag.$^{*}$ &Reduced mag.&Filter &Observatory\\
\hline
 (468861) 2013 LU28& 2019 08 01.80 &13.06&13.29&4.3&21.02$\pm$0.06      &9.82   &B&CrAO\\ 
                             &&&&& 20.18$\pm$0.04       &8.98   & V &  CrAO\\
                            &&&&& 19.56$\pm$0.08                &8.36   & R &  CrAO\\
                            &&&&& 19.46$\pm$0.09                &8.26   & I &  CrAO\\
    & 2019 08 31.77 & 12.95 & 13.42 & 3.9 & 20.92$\pm$0.06      &9.72   & B &  CrAO\\
                                      &&&&& 20.09$\pm$0.04      &8.89   & V &  CrAO\\
                                      &&&&& 19.52$\pm$0.08      &8.32& R &  CrAO\\
                                      &&&&& 18.85$\pm$0.09      &7.65   & I &  CrAO\\
     & 2020 03 22.05 & 12.20 & 11.66 & 4.0 & 20.64$\pm$0.12     &9.44   & B & Terskol\\
                                       &&&&& 19.72$\pm$0.02     &8.95   & V & Terskol\\
                                      &&&&& 19.20 $\pm$0.02     &8.43   & R & Terskol\\
(517717) 2015 KZ120 & 2019 08 01.84 &8.41&8.26&6.9&20.83$\pm$0.03       &11.62  & B & CrAO\\
                        &&&&& 19.89$\pm$0.03            &10.68  & V &  CrAO\\
                                  &&&&& 19.47$\pm$0.04  &10.26  & R &  CrAO\\
                                  &&&&& 18.92$\pm$0.06  &9.71   & I &  CrAO\\
  & 2019 08 30.81 & 8.42 & 8.39 & 6.9 & 20.74$\pm$0.04  &11.49  & B &  CrAO\\
                                  &&&&& 19.96$\pm$0.01  &10.71  & V &  CrAO\\
                                  &&&&& 19.47$\pm$0.02  &10.22  & R &  CrAO\\
                                  &&&&& 18.85$\pm$0.05  &9.60   & I &  CrAO\\                                     
2020 EP & 2020 04 23.97 &2.55 & 1.78 & 17.6 & 20.95$\pm$0.05 &17.67     & B &  CrAO\\
                            &&&&& 20.12$\pm$0.05                &16.84  & V &  CrAO\\
                            &&&&& 19.61$\pm$0.04                &16.33  & R &  CrAO\\
                            &&&&& 19.19$\pm$0.09                &15.91  & I &  CrAO\\
A/2019 U5 & 2020 07 19.91 &8.67&8.32&6.4& 19.79$\pm$0.05        &10.50  & B & CrAO\\
                                &&&&&   19.00$\pm$0.05  &9.71   & V &  CrAO\\
                                &&&&&   18.59$\pm$0.05  &9.30   & R &  CrAO\\
                                &&&&&   18.04$\pm$0.05          &8.75   & I &  CrAO\\ 
 & 2020 07 27.96 & 8.64 & 8.28 & 6.4 &   19.73$\pm$0.04         &10.46  & B & Terskol\\
                                &&&&&   19.05$\pm$0.04  &9.78   & V & Terskol\\
                                &&&&&   18.63$\pm$0.04  &9.36   & R & Terskol\\
                                &&&&&   18.12$\pm$0.04          &8.85   & I & Terskol\\
 & 2020 08 14.94 & 8.53 & 8.23 & 6.6 & 19.74$\pm$0.04   &10.51  & B & CrAO\\
                         &  & &  &  & 18.95$\pm$0.03    &9.72   & V & CrAO\\ 
                        &  &  & & & 18.52$\pm$0.03      &9.29   & R & CrAO\\ 
                     &  & &  &  & 17.97$\pm$0.03                &8.74   & I & CrAO\\ 
C/2018 DO4 &2019 10 01.05 &2.45&1.99&23.4& 14.05$\pm$0.08       &10.61  & B &  CrAO\\
                                     &&&&& 13.13$\pm$0.03       &9.69   & V &  CrAO\\
                                     &&&&& 12.70$\pm$0.03       &9.26   & R &  CrAO\\
                                     &&&&& 11.99$\pm$0.07       &8.55   & I &  CrAO\\
 & 2019 10 03.06 & 2.45 & 1.97 & 23.0 & 13.34$\pm$0.04  &9.92   & R &  Terskol\\
 & 2019 10 05.08 & 2.46 & 1.93 & 22.5 & 14.22 $\pm$0.06         &10.84  & B &  Terskol\\
                                  && &&& 13.32 $\pm$0.04        &9.94   & V & Terskol\\
                                   &&&&& 12.88 $\pm$0.04        &9.50   & R & Terskol\\                             
C/2019 O3 & 2020 07 19.92 &8.92&8.27&5.2&  18.30$\pm$0.03       &8.71   & B &  CrAO\\
                            &&&&&  17.54$\pm$0.03       &7.95   & V &  CrAO\\
                            &&&&&   17.11$\pm$0.03      &7.52   & R &  CrAO\\
                            &&&&&   16.69 $\pm$0.09     &7.10   & I &  CrAO\\                     
& 2020 08 14.90 &8.90&8.32&5.5&  18.28$\pm$0.03         &8.93   & B &  CrAO\\
                     &&&&&  17.54$\pm$0.03              &8.19   & V &  CrAO\\
                    &&&&&  17.14$\pm$0.03               &7.79   & R &  CrAO\\
                    &&&&&  16.76$\pm$0.03               &7.41   & I &  CrAO\\

        \hline
    \end{tabular}
    
\begin{flushleft}
        $^{*}$ The magnitudes of C/2018 DO4 and C/2019 O3 were calculated for the 10~000~km and 15~000~km apertures, respectively. 
\end{flushleft}    

\end{table*}

\subsection{(468861) 2013 LU28}

(468861) 2013~LU28 (previously designated as 2014~LJ9 and 2015~KB157) is a TNO on a retrograde orbit ($i=125.4^\circ$) with a very high eccentricity $e=0.95$.
The semi-major axis of its orbit is $a=187.1$~au, and perihelion and aphelion distances are $q=8.7$~au and $Q=365.5$~au, respectively.

The photometric observations were carried out in BVRI filters at CrAO during two nights
in August 2019 and in BVR filters at Terskol observatory on March 21, 2020.
No colour variation was found for the different dates.
The obtained average surface colours ($B-V = 0.83\pm0.04$, $V-R=0.57\pm0.04$, and $R-I = 0.68\pm0.05$) imply a moderately red surface of 2013~LU28.
The differences in reduced magnitude values measured at different dates are within the estimated brightness errors.

The absolute magnitude was calculated using the formalism by \citet{Bowell1989} and assuming $G=0.15$.
Then the visible absolute magnitude for 2013~LU28 is $H_{V}=8.6\pm0.1$~mag, 
which is fainter than the value adopted in MPC (8.2~mag).
In order to estimate the diameter of the object, we assume a low to moderate 
albedo in 0.04-0.1 range. In this case the diameter is 80-127~km.

\subsection{(517717) 2015~KZ120}

2015 KZ120 is a Centaur on a near-polar ($i=85^\circ$) and highly eccentric ($e=0.82$)
orbit with $a=46.7$~au, $q=8.4$~au, and $Q=85.0$~au.

Our observations were carried out during two nights in August 2019 at CrAO in BVRI photometric filters.
The measured surface colours $B-V = 0.78\pm0.05$, $V-R=0.49\pm0.03$, and $R-I = 0.61\pm0.06$
suggest that the object has a moderately red surface.

No significant variation is found in the values of reduced magnitude.
The calculated visible absolute magnitude is $H_{V}=10.2\pm0.1$~mag, which is similar 
to that reported in MPC (10.0~mag). The estimated diameter (again assuming the albedo in 0.04-0.1 range) is 39-61~km.

\subsection{2020 EP}
2020 EP is a recently discovered object on a near-polar ($i=76^\circ$)
and very eccentric orbit ($e=0.76$). Its semi-major axis is $a=10.5$~au, 
the perihelion distance is $q=2.5$~au, and the aphelion distance $Q=18.5$~au.

2020 EP was observed during one night on April 23, 2020, at CrAO observatory in BVRI filters.
The object is on a cometary orbit and at the time of observation
it was close to its perihelion at a heliocentric distance $r=2.55$~au.
However, we found no signs of cometary activity for 2020~EP.
The measured surface colours correspond to a moderately red surface
($B-V = 0.83\pm0.09$, $V-R=0.51\pm0.06$, and $R-I = 0.42\pm0.10$).

The calculated visible absolute magnitude is $H_{V}=15.9\pm0.1$~mag and is similar to the value in MPC (16.0~mag). 
Using this value of absolute magnitude, the estimated diameter of the object is rather small, only 3-4~km.

\subsection{A/2019 U5 (A/PanSTARRS)}

A/2019~U5 (A/PanSTARRS) (hereafter A/2019~U5) is a recently discovered object on a hyperbolic cometary orbit with $i=113.5^\circ$, $e=1.002$, and $q=3.6$~au.
With the current orbital elements A/2019~U5 will be reaching its perihelion in March 2023.
To date, no cometary activity has been detected for this object, thus it is currently classified as a hyperbolic asteroid. 
Very few such asteroids have been discovered to date, and the majority of them eventually showed signs of cometary activity.

We observed A/2019~U5 during three nights in the BVRI filters. 
On July 19 and August 14, 2020, the object was observed at CrAO, and on July 27, 2020, at Terskol observatory.
The obtained colour indices are similar and correspond to a moderately red surface. 
The average colours for three nights are $B-V = 0.73\pm0.06$, $V-R=0.42\pm0.06$, and $R-I = 0.55\pm0.06$.
The calculated reduced magnitudes are very similar for all of our observations.

The estimated absolute magnitude is $H_{V}=9.3\pm0.1$~mag (compared to $H_{V}=9.68\pm0.68$ in the JPL
Small-Body Database\footnote{https://ssd.jpl.nasa.gov/sbdb.cgi}), and the corresponding diameter is 60-95~km.

\subsection{C/2018 DO4 (Lemmon)}

C/2018 DO4 (Lemmon) (hereafter C/2018 DO4) is a retrograde comet ($i=160^\circ$) with $e = 0.91$, $a = 25.9$~au, $q = 2.41$~au, and $Q = 49.3$~au. 
The comet reached  perihelion on August 18, 2018 and was observed post-perihelion at CrAO on October 02, 2019 and at the Terskol observatory on October 03 and 05, 2019.
A cometary coma is clearly seen in all of the obtained images.

\begin{figure}
        \includegraphics[width=\columnwidth]{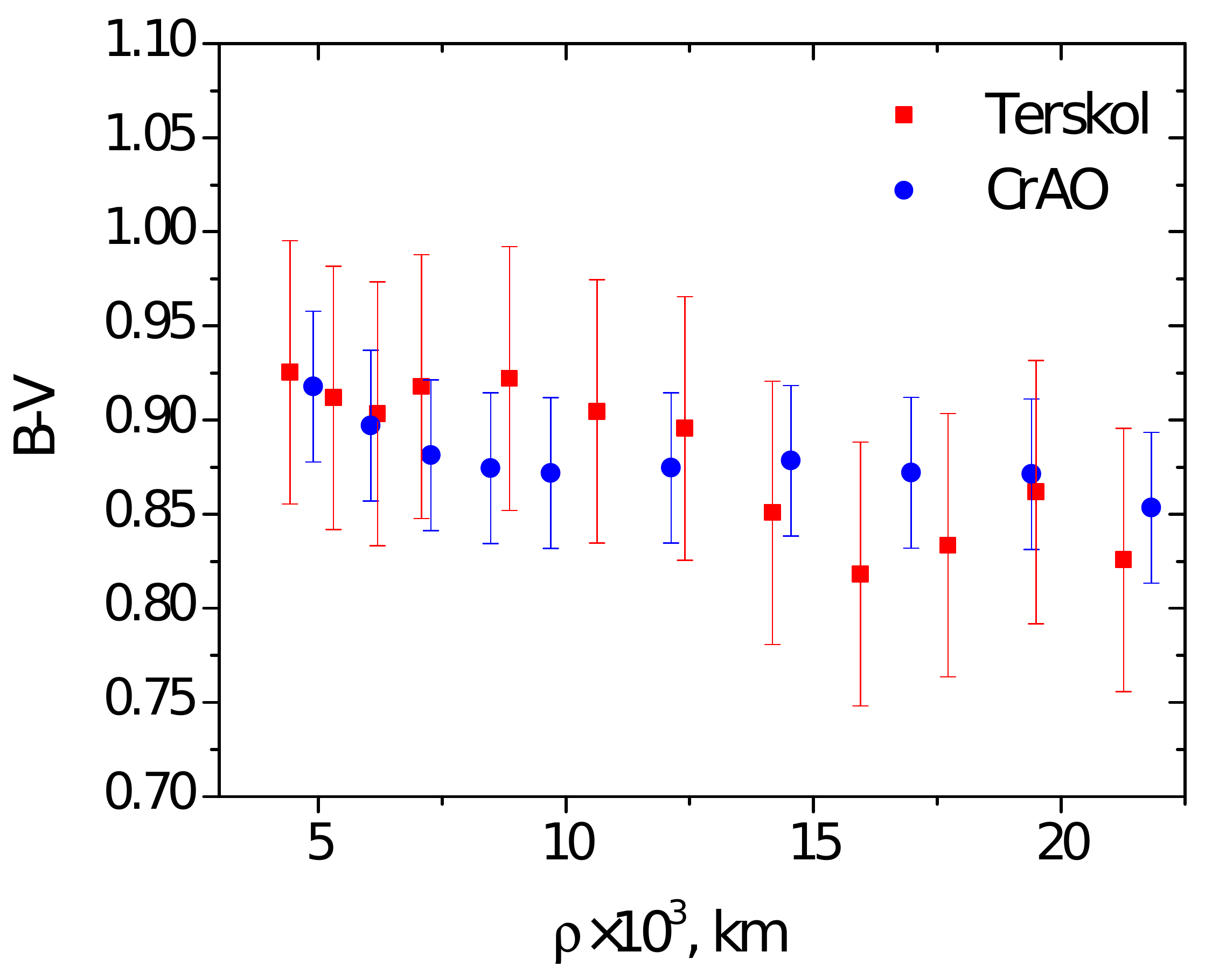}
        \caption{B-V colours of the comet C/2018 DO4 at different aperture sizes.}
        \label{color_grad}
\end{figure}

Colours of the comet were calculated using different apertures that correspond to a linear size range of about $\sim$5~000-20~000~km. 
The obtained colours are shown in Table~\ref{comet_table}. 
Additionally, in Fig.~\ref{color_grad} we present the B-V colours that were obtained at different apertures.
 No significant changes of colours are detected in the range from $\sim$5~000 to 20~000~km. 

\begin{table}
    \centering
    \caption{Colours of the comet C/2018 DO4 obtained on October 03 and 05, 2019.}
        \label{comet_table}
        \begin{tabular}{lllll}
        \hline
        \hline
Date&$\rho$, km&B-V&V-R&V-I\\
        \hline
Oct 02&$\sim$5 000 &0.93$\pm$0.04&0.42$\pm$0.04&1.14$\pm$0.08\\
&$\sim$10 000 &0.92$\pm$0.04&0.43$\pm$0.04&1.14$\pm$0.08\\
&$\sim$15 000 &0.89$\pm$0.04&0.43$\pm$0.04&1.15$\pm$0.08\\
&$\sim$20 000 &0.90$\pm$0.04&0.47$\pm$0.04&1.27$\pm$0.08\\

Oct 05&$\sim$5~000&0.91$\pm$0.07&0.44$\pm$0.05&\\
&$\sim$10 000&0.92$\pm$0.07&0.41$\pm$0.05&\\
&$\sim$15 000&0.85$\pm$0.07&0.44$\pm$0.05&\\
&$\sim$20 000&0.86$\pm$0.07&0.45$\pm$0.05&\\
        \hline
    \end{tabular}
\end{table}

\begin{figure}
        \includegraphics[width=\columnwidth]{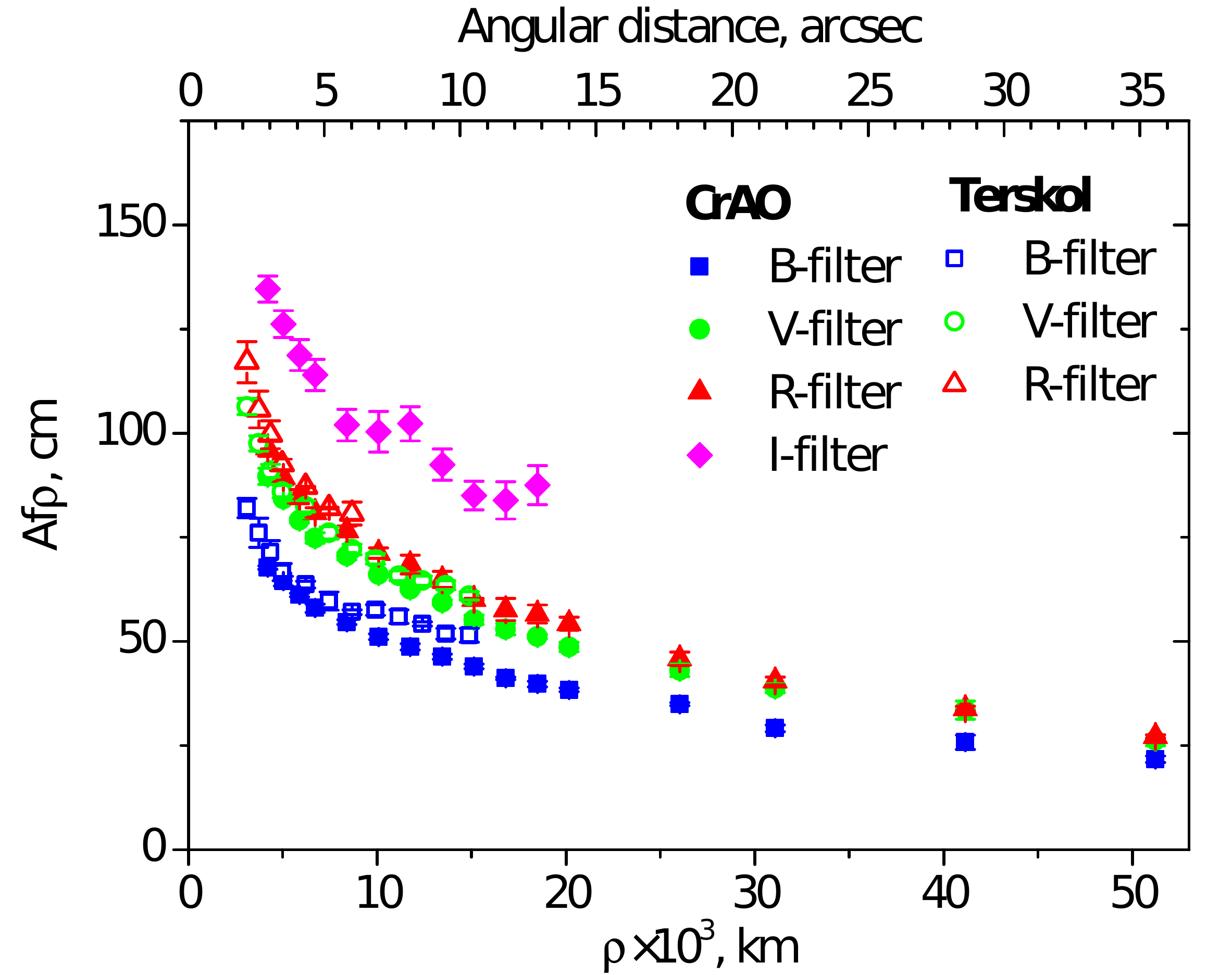}
        \caption{Values of \textit{Af$\rho$} for C/2018 DO4 (Lemmon) at different apertures. 
        The top x-axis shows the angular aperture size.
        The bottom x-axis shows the linear aperture size.
        The filled symbols correspond to observations at CrAO made on October 01 and the open symbols correspond to observations at the Terskol observatory made on October 03.}
        \label{afrho_all}
\end{figure}

The level of dust production for a comet can be quantified by means of the parameter \textit{Af$\rho$} \citep{A'Hearn84}.
It contains the value of the geometric albedo \textit{A}; the radius of the photometric aperture $\rho$ in cm; and the factor \textit{f}, which is the total scattering cross-section of the grains within this aperture.
This parameter can be estimated as
\begin{equation}
Af\rho = \frac{4r^{2}\Delta^{2}10^{0.4(m_{sun}-m_{c})}}{\rho},
\end{equation}
where $\Delta$ is the geocentric distance in au, $r$ is the heliocentric distance in cm, and 
$m_{sun}$ and $m_{c}$ are the apparent magnitudes of the Sun and the comet, respectively.
We used the solar colours from \citet{Holmberg2006}.
In Fig.~\ref{afrho_all} we show the values of \textit{Af$\rho$} that were calculated for different filters on October 02 and 05, 2019.
The decrease in \textit{Af$\rho$} with the distance from the nucleus most likely can be explained by the fragmentation of the particles as the fragmentation causes a decrease in the particle scattering cross section.
We see a noticeable difference in the radial profiles for the B and I filters. 
Since the light scattering characteristics of particles change quickly when the particle size is comparable with the wavelength, this confirms the predominance of submicron- and micron-sized particles in the cometary coma. 
The values of \textit{Af$\rho$} in the R filter for the 6~000~km aperture is 97$\pm$3~cm and 103$\pm$5~cm for observations on October 02 and 05, 2019, respectively, which suggest that the level of dust production did not change significantly.

In order to reveal low-contrast structures in the coma of comet C/2018~DO4 the images were treated with digital filters. 
We applied a combination of four numerical techniques: rotational gradient   \citep{Larson1984}, 1/$\rho$ profile, azimuthal average, and renormalization  \citep{Samarasinha2014}. 
The first two techniques allow us to remove the bright background from a cometary coma and to highlight the low-contrast features. 
Division by the azimuthal average is very good at separating brighter broad jets. 
We applied each technique to the individual frames and to the composite images in order to evaluate whether the revealed features are real or not. 
We also studied the change in the jet structure due to the shift of the comet's optocentre. Even in the case of significant displacement of the optocentre by 6 pixels, the detected structures were preserved for all images.
Previously this technique was used to pick out structures in several comets with good results \citep[e.g.][]{Ivanova2018, Ivanova2019, Picazzio2019}.

\begin{figure*}
        \includegraphics[width=\textwidth]{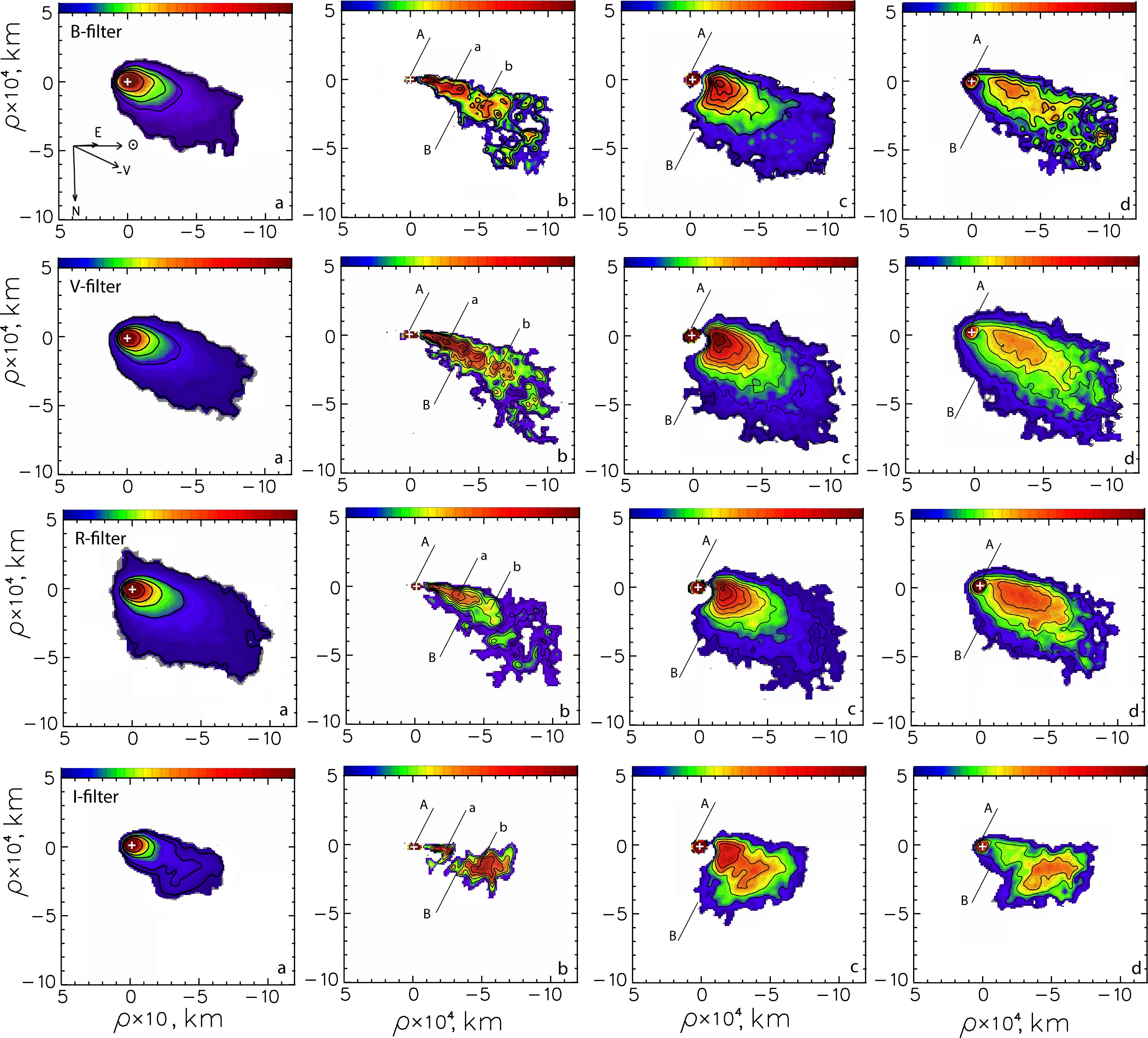}
        \caption{Relative intensity images of the comet C/2018 DO4 (Lemmon) in the B, V, R, and I
        filters obtained on October 02, 2019, at CrAO. 
        (a) Direct images of the comet with the isophots differing by a factor $\sqrt{2}$; 
        (b)  Image processed by a rotational gradient method \citep{Larson1984}; 
        (c) and (d) Images to which the azimuthal average and the division by 1/$\rho$
        profile methods were applied, respectively \citep{Samarasinha2014}. 
        The arrows show the direction to the Sun ($\odot$),
        north (N), east (E), and the negative velocity vector of the comet projected
        onto the sky plane (V). Negative distance is in the solar direction.}
        \label{morph_CrAO}
\end{figure*}

\begin{figure*}
        \includegraphics[width=\textwidth]{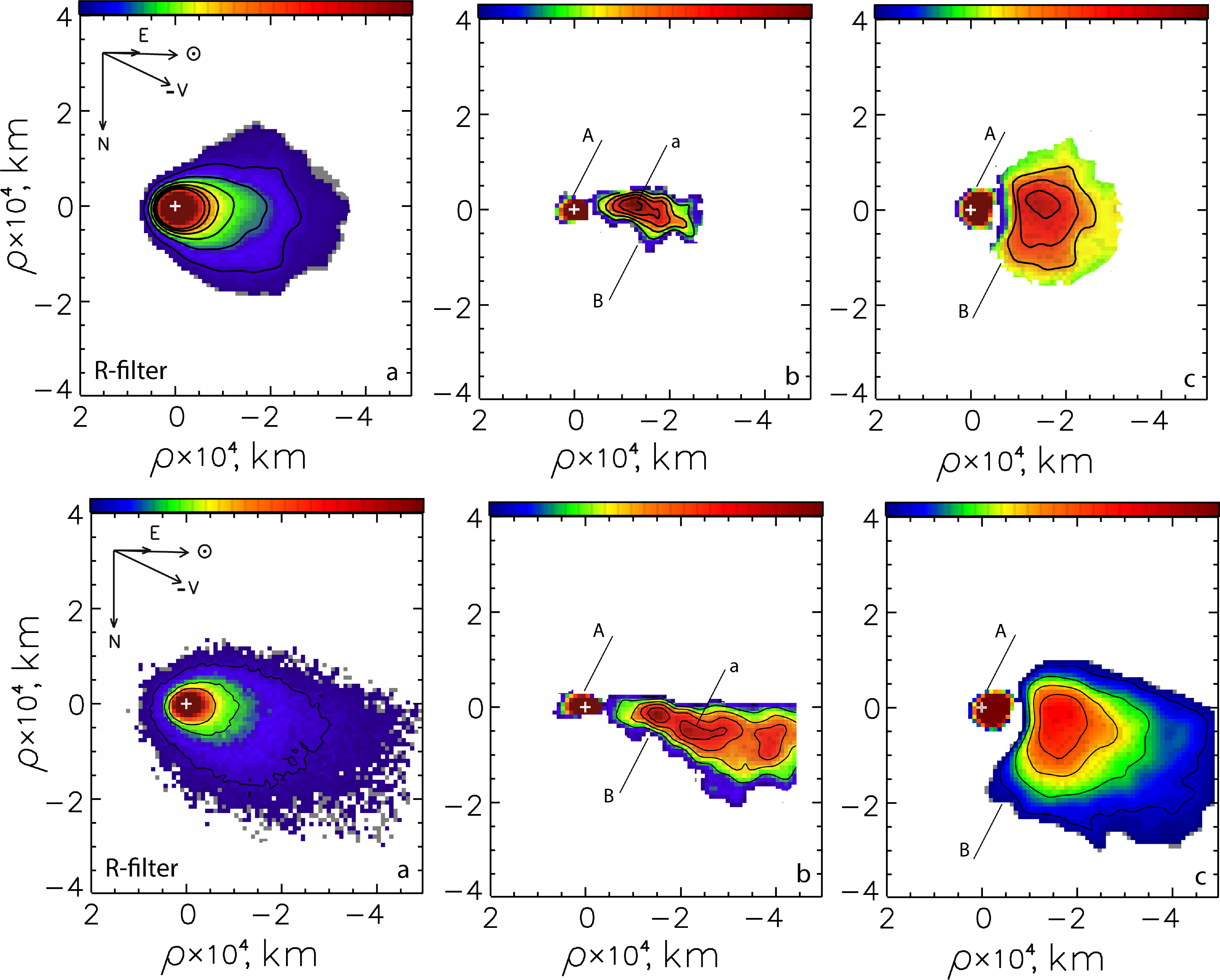}
        \caption{Relative intensity images of the comet C/2018 DO4 (Lemmon) in the
        R filter obtained on October 03, 2019 (top), 
        and on October 05, 2019 (bottom), at Terskol Observatory. 
        (a) Direct images of the comet with the isophots differing by a factor $\sqrt{2}$; 
        (b) Image processed by the rotational gradient method \citep{Larson1984}; 
        (c) Image to which the azimuthal average method was applied \citep{Samarasinha2014}.
        The arrows show the directions to the Sun ($\odot$), north (N), east (E),
        and the negative velocity vector of the comet projected onto the sky plane (V).
        Negative distance is in the solar direction.}
        \label{morph_Terskol}
\end{figure*}

Figure~\ref{morph_CrAO} shows the direct B, V, R, and I images of the comet in relative intensity obtained on October 02, 2019. 
Figure~\ref{morph_Terskol} shows the direct R images of the comet obtained on October 02 and 05, 2019. The relative intensities of the adjacent contours of the isophots differ by a factor of $\sqrt{2}$. 
The observed coma is strongly asymmetric, and most probably was formed due to activity of an isolated area on the cometary nucleus.
In processed images we revealed two strong structures: a bright near nucleus region (A) and a tail (B), which is elongated along the negative velocity vector in the direction of the coma.
The position angle of the tail, measured anticlockwise from the north through the east, is $75^{\circ}$. 

In CrAO images we can see two isolated structures ($a$, $b$) placed along the tail at distances of 20~000--60~000~km from the optocentre (Fig.~\ref{morph_CrAO});
structure $a$ can also be  seen in the Terskol images (Fig.~\ref{morph_Terskol}). 
These structures are most pronounced in the images processed by the rotational gradient method because this method is very sensitive to faint structures in cometary coma. 
The rotational gradient technique highlights the structures that are distributed radially from the nucleus. In general, these structures are formed by the effects of the nucleus rotation on the released particles.
The separated structures were observed for all observation sets from October 02 to 05, 2019. 
Following \citet{Farnham2009}, we define these features as arcs or shells. 
Most likely the visible separation into such shells is a result of the nucleus rotation and the projection of a particle flow that was ejected from the nucleus.
The collimated appearance of a tail and structures, which are similar in all the filters that were used, favours the dust-dominated nature of the comet. 
A more detailed analysis and modelling of the morphology of the comet   will be addressed in a future work.

\subsection{C/2019 O3 (Palomar)}

C/2019 O3 (Palomar) (hereafter C/2019 O3) is another object on a near-polar ($i=89.8^\circ$) hyperbolic ($e=1.002$) orbit with $q=8.8$~au.
It is classified as a comet as  cometary activity has been detected. 
The object will reach its perihelion in March 2021.

We observed the comet C/2019 O3 during two nights on July 19 and August 14, 2020, at CrAO in the BVRI filters.
The calculated colours that correspond to the aperture size of 15~000~km are
$B-V = 0.75\pm0.03$, $V-R=0.47\pm0.03$, and $V-I = 0.94\pm0.09$.
Table~\ref{comet2_table} contains the surface colours that were measured at different apertures.

\begin{table}
    \centering
    \caption{Colours of the comet C/2019 O3 obtained on July 19 and August 14, 2020.}
        \label{comet2_table}
        \begin{tabular}{lllll}
        \hline
        \hline
Date &$\rho$, km&B-V&V-R&V-I\\
        \hline
Jul 19 & $\sim$30 000 &0.76$\pm$0.03&0.44$\pm$0.04&0.85$\pm$0.08\\
       & $\sim$40 000 &0.80$\pm$0.03&0.41$\pm$0.04&0.85$\pm$0.09\\
        &$\sim$50 000 &0.81$\pm$0.03&0.43$\pm$0.04&085$\pm$0.09\\
        &$\sim$60 000 &0.78$\pm$0.03&0.43$\pm$0.04&0.95$\pm$0.09\\
        
Aug 14& $\sim$30 000 &0.74$\pm$0.03&0.41$\pm$0.04&0.78$\pm$0.08\\
       & $\sim$40 000 &0.72$\pm$0.03&0.45$\pm$0.04&0.82$\pm$0.09\\
        &$\sim$50 000 &0.73$\pm$0.03&0.43$\pm$0.04&086$\pm$0.09\\
        &$\sim$60 000 &0.72$\pm$0.03&0.43$\pm$0.04&0.93$\pm$0.09\\
        \hline
    \end{tabular}
\end{table}

\begin{figure}
        \includegraphics[width=\columnwidth]{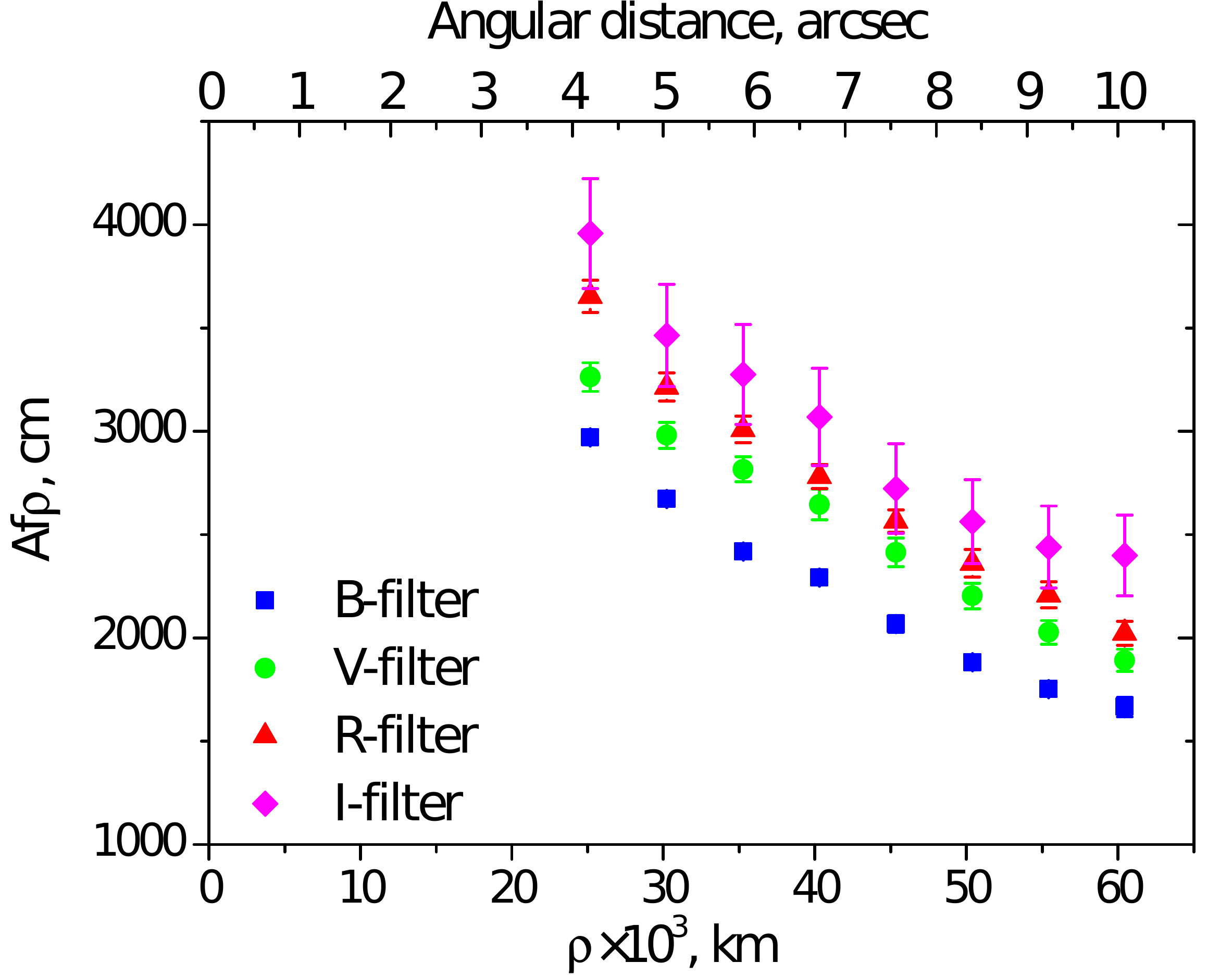}
        \caption{Average values of the parameter \textit{Af$\rho$} for the comet C/2019 O3  observed on July 19 and August 14, 2020.
        The top x-axis shows the angular aperture size in arcseconds and the bottom x-axis shows the linear aperture size (in km).}
        \label{afrho_2019O3}
\end{figure}

The calculated value of \textit{Af$\rho$} for observations in July and August, 2020, are similar within the uncertainties, thus we provide here the average values. 
In the R filter and for the aperture size of about 25~000~km the value of $Af{\rho}$ is $3654\pm77$~cm, 
which is significantly larger than that for the comet C/2018~DO4.
In Fig.~\ref{afrho_2019O3} we show the mean values of $Af\rho$ in the BVRI filters calculated with various apertures.

As for the comet C/2018 DO4 we obtained a noticeable difference in the \textit{Af$\rho$} profiles for the B and I filters. Again, this confirms the predominance of submicron- and micron-sized particles in the cometary coma of C/2019 O3. 

The cometary activity was detected in our images when the object was at $r=8.9$~au from the Sun.
This implies that the activity mechanism must be different from water ice sublimation, which 
is only possible at a distance from the Sun of up to $3-4$~au  \citep{Meech2004}. 
At larger distances other mechanisms are considered, such as
sublimation of CO and/or CO${_2}$, transition between amorphous and crystalline phases of water ice, or annealing of the amorphous water ice \citep{Meech2009, Prialnik1992, Luu1993}.

\section{Search for high-inclination objects in multi-band sky surveys}

In addition to our observations, we performed a search for high-inclination objects
in databases. In particular, we used data from the Sloan Digital Sky Survey (SDSS) and Pan-STARRS surveys.
For the SDSS we used data from a catalogue and for the Pan-STARRS we worked directly with the images. 
 The SDSS and Pan-STARRS surveys both used Sloan photometric filters.
Thus, in order to compare the obtained colours with those in the Johnson--Cousins
photometric system we used the transformation equations from \citet{Kostov18}.

\subsection{SDSS catalogue}

The Sloan survey provides astrometric and photometric observations of a large variety of astronomical objects, including small solar system bodies.
The photometric observations were conducted using the 2.5m wide-angle telescope at the Apache Point Observatory (New Mexico, United States) \citep{York2000}.
The camera consists of 30 2048$\times$2048 photometric CCDs that are mosaicked on the telescope's focal plane, which allows a field of view of $3^{\circ}$ \citep{Gunn1998}.

The SDSS camera obtained quasi-simultaneous ugriz images with a time interval of approximately 72~s between each exposure,
and the largest temporal baseline is about  300s between the r and g filters.
The selection of retrograde asteroids was carried out from the SDSS SSO catalogue \citep{Sergeyev2020},
which contains more that 1M measurements of solar system objects observed in 1998-2009.
It was prepared by analysing the moving objects that were extracted from the DR16 and the  Stripe82 SDSS photometry catalogues. 
To distinguish known asteroids, the cone-search queries of SkyBoT service were used \citep{Berthier2006,2016-MNRAS-458-Berthier}.

We found only four high-inclination objects with multiple observations:
Centaur 2005 VD, TNOs 2005 OE and 2012 DR30, and a NEO 2012 MS4. 
Given the rather large magnitude errors for the individual objects, we consider here only the average colours of the found objects:  B-V=0.86$\pm$0.18, V-R=0.53$\pm$0.11, and R-I=0.40$\pm$0.15.
Thus, although the use of SDSS data did not help to increase the statistics on colours of highly inclined objects,
we can conclude that the calculated mean colours are consistent with moderately red surfaces.

\subsection{Pan-STARRS images}

We used the data obtained by the Pan-STARRS survey telescope \citep{Tonry12}
during the first 18 months from March 2010 to May 2014.
The Pan-STARRS survey was conducted using the 1.8m 
telescope of the Haleakala observatory (Hawaii, United States). 
The telescope is equipped with a 1.4 gigapixel CCD camera GPC1 that covers a field of view of $\sim7\times7^{\circ}$.
The grizy filter system used by the survey is similar to the Sloan photometric system \citep{Tonry12}.
In this work we only used the data obtained in the g, r, and i filters.
The search for the objects that are present on Pan-STARRS images
was performed using the Solar System Object Image Search \citep{Gwyn12}.

We  found quasi-simultaneous observations for three TNOs: (336756) 2010 NV1, (342842) 2008 YB3, and 2012 DR30.
The measured colours for all found TNOs correspond to moderately red surfaces 
consistent with the previously measured colours for these objects (see Table~\ref{Result table}).
Similarly to SDSS, we found very few high-inclination objects in the Pan-STARRS images.
It appears that the study of these rare and faint objects using all-sky survey data is not 
very fruitful, and that a dedicated programme is necessary.

\section{Analysis of surface colours for high-inclination asteroids}

Our new observations of four high-inclination asteroids showed that all of these extreme objects are moderately red.
Similarly, the colours of several other high-inclination objects that were found in the databases suggest that they also have moderately red surfaces.
In Table~\ref{Result table} we collect all the available colours for high-inclination objects. 
Previously the colour indices were measured for 21 high-inclination asteroids. 
Our new observations increase the number of objects with known colours, 
but the number of studied high-inclination objects remains very small (around 10\% of the discovered objects).
The table contains orbital elements ($i, a, e$); albedos and diameters,
if available; absolute magnitudes from MPC; and surface colours in the Johnson--Cousins photometric system. 
As   can be seen, all of the studied high-inclination objects have surface
colours from neutral to moderately red, which confirms the conclusion by \citet{Jewitt15} 
on the absence of extremely red surfaces among these objects.  

\begin{figure}
        \includegraphics[width=\columnwidth]{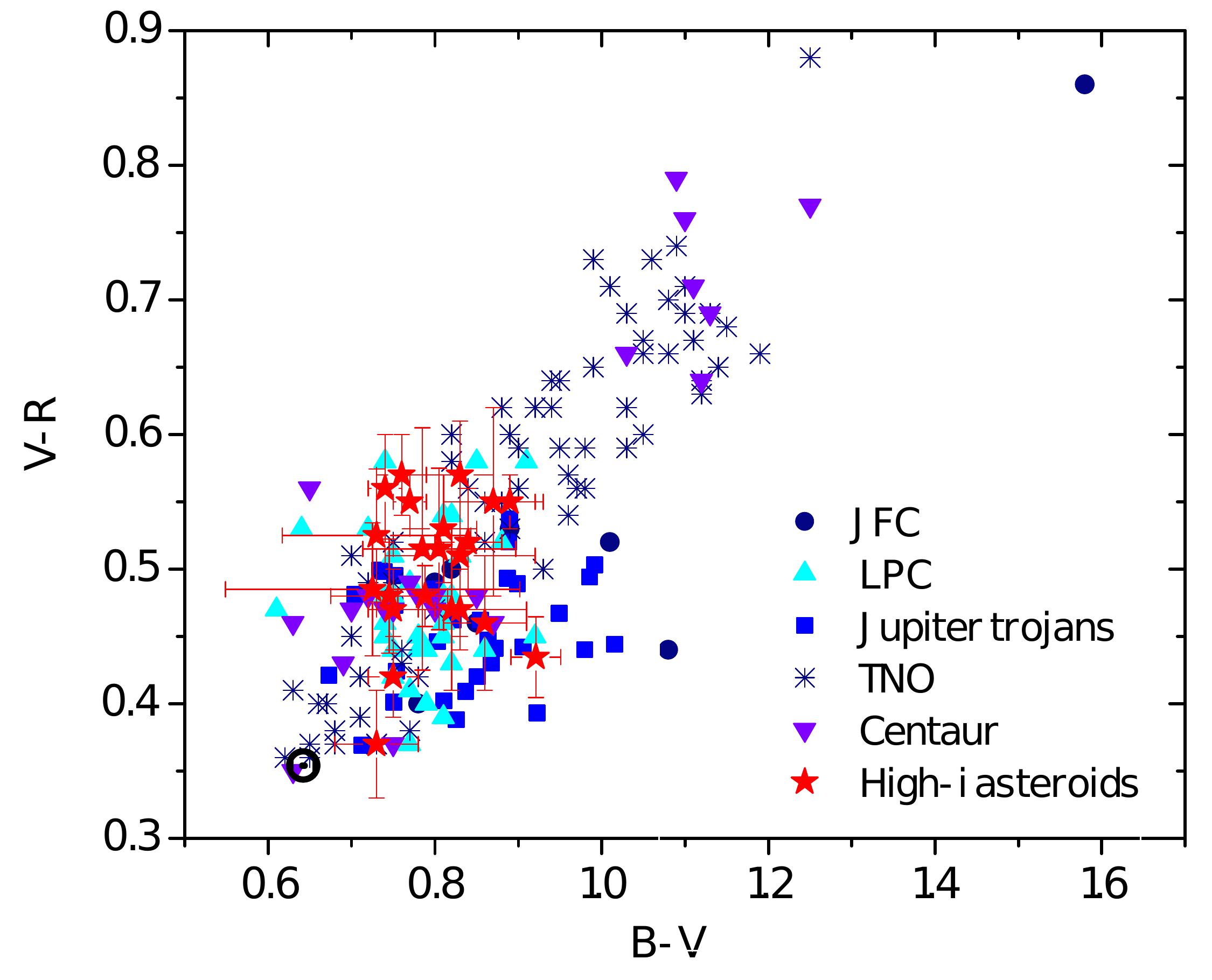}
        \caption{Colour-colour diagram comparing different dynamical classes of small solar system objects, such as Jupiter family comets (JFC) \citep{Lamy2009},
        long-period comets (LPC) \citep{Solontoi2012}, Jupiter trojans \citep{Emery2011}, 
        TNOs, and Centaurs \citep{Belskaya2015} with objects on highly inclined orbits. The black symbol shows the solar colours.}
        \label{color-color}
\end{figure}

Such surface colours are inherent for different dynamical classes of solar system objects.
Figure~\ref{color-color} shows the colour-colour diagram of high-inclination objects together
with other dynamical types, such as Jupiter trojans, Centaurs, TNOs, and comets.
As   can be seen, the high-inclination objects (including those that were observed in this work)
do not show a wide variety of colours, unlike TNOs or Centaurs. 
They fall in a rather restricted area intersecting with grey Centaurs, moderately red TNOs, and Jupiter trojans.

\begin{figure}
        \includegraphics[width=\columnwidth]{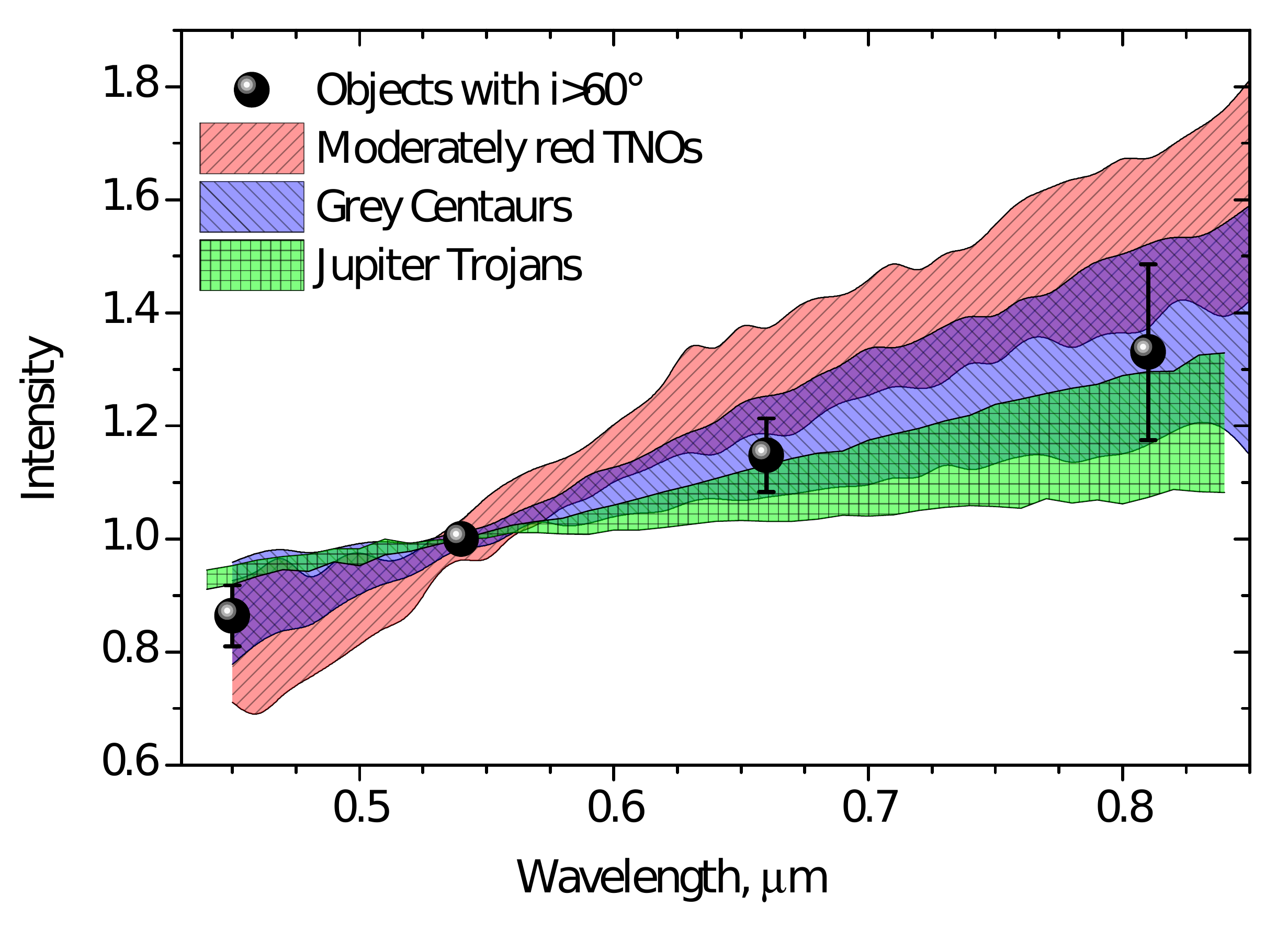}
        \caption{Mean colours of high-inclination objects (black circles) 
        together with mean spectra of Jupiter trojans (green) from
        \citet{Emery2011}, grey Centaurs (blue), and moderately 
        red TNOs (red) from \citet{Merlin2017}.}
        \label{spec-fill}
\end{figure}

To further explore the similarity between these populations, we compare the average spectra
of  trojans \citep{Emery2011}, grey Centaurs, and moderately red TNOs \citep{Merlin2017} with that of high-inclination objects.
To do this we converted the average colours and the corresponding standard
deviation into intensities and subtracted the Sun colours (Fig.~\ref{spec-fill}).
The mean colours of high-inclination objects have the best match with grey  Centaurs,
but also overlap with those of moderately red TNOs and to a lesser degree with those of Jupiter trojans.

There is a tendency for ultra-red material to be mostly present among low-inclination objects \citep{Tegler2000, Trujillo2002,Doressoundiram2005,Gulbis2006,Fulchignoni2008,Peixinho2008}.  
As Fig.~\ref{color-color} shows, there is a clear absence of very red surfaces
among polar and retrograde objects, even though they are present on some TNOs and Centaurs.
Different scenarios are proposed in order to explain this absence in comets
\citep{Jewitt2002} and damocloids \citet{Jewitt05,Jewitt15}.
One is the size-dependent effects of collisional resurfacing as measured damocloids
are smaller  than     Centaurs and KBOs. 
This scenario was considered the least plausible by \citet{Jewitt05}.
It is not supported by the new observations of relatively
large high-inclination objects (or damocloids), as can be seen in Table~\ref{Result table}.

Another explanation is the thermal dissociation or the instability of ultra-red   matter
at higher temperatures \citet{Jewitt05,Lamy2009}. 
It is suggested that higher temperatures are uncharacteristic
for the complex organics   associated with the ultra-red matter \citep{DalleOre2015, Jewitt2002}.
What somewhat decreases the likelihood of such scenario is the discovery of very red 
objects that come very close to the Sun and retain their very red surfaces.
For example, the objects on cometary orbit 2016~ND21 ($q=3.76$~au, $e=0.56$, $a=8.46$~au)
was observed at heliocentric distance $r=3.8$~au and was found to have an extremely red surface
colours similar to those of very red TNOs and Centaurs \citep{Hromakina2019}.
However, this object's orbit is highly eccentric, and the body spends 
a lot of time in the outer part of the solar system, and may not have spent enough
time at close distances to lose its redness.

Finally, the blanketing of the primordial red surfaces created by outgassing activity may be responsible for the absence of ultra-red surfaces \citep{Jewitt2002, Grundy2009}.
This scenario, although very plausible, has  unresolved issues. 
In particular, the discovery of a wide variety of colours among active comets, including
ultra-red ones \citep{Lamy2009, Bauer2003}, requires a better understanding of the blanketing mechanism.  
It appears that the existence of activity does not necessarily result in the creation of a mantle that covers the ultra-red matter on the surface \citep{Lamy2009, Bauer2003},
and other effects may prevent the dust from settling on the surface.
It is possible that such the mantle would appear only after a prolonged activity.

\begin{table*}
\small
    \centering
    \caption{Orbital and physical properties of high-inclination objects}
        \label{Result table}
        \begin{tabular}{lllllllllll}
            \hline
            \hline
Object&i&  e & a, au & H$^1$, mag & A$^2$ & D, km & B-V&V-R&R-I & Ref$^3$\\
\hline
(330759) 2008 SO218 &170.3&0.56&8.12&12.8&0.08&13.5&0.89$\pm$0.03&0.55$\pm$0.02&0.50$\pm$0.04&J15\\
(336756) 2010 NV1 &140.7&0.97&281.2&10.4&   0.04&52.2&0.79$\pm$0.03&0.53$\pm$0.02&0.39$\pm$0.02&J15\\
                     &&&&&&&0.74$\pm$0.03&0.50$\pm$0.02&&T16\\
                     &&&&&&&0.87$\pm$0.10&0.53$\pm$0.10&& \textbf{This work}\\
            
(342842) 2008 YB3  &105.1&0.44&11.62&9.3&0.05&79.0&0.82$\pm$0.01&0.46$\pm$0.01&&S10\\
                                &&&&&&&&0.50$\pm$0.06&0.5$\pm$0.08&P13\\
                                &&&&&&&0.80$\pm$0.01&0.46$\pm$0.01&0.95$\pm$0.01&H12\\
                &&&&&&&0.74$\pm$0.02&0.51$\pm$0.02&&J15\\
                                &&&&&&&0.77$\pm$0.02&0.46$\pm$0.02&&T16\\
                                &&&&&&&0.81$\pm$0.06&0.49$\pm$0.06&&   \textbf{This work}\\

(418993) 2009 MS9 &68.0&0.97&377.9 &9.8&&&0.84$\pm$0.04&0.52$\pm$0.04&&J15\\

(434620) 2005 VD &172.9&0.25&6.67&  14.1&&&0.60$\pm$0.17&0.45$\pm$0.11&&P13\\ 

(468861) 2013 LU28 &125.4&0.95&187.10&8.2&&& 0.83$\pm$0.04&0.57$\pm$0.04&0.68$\pm$0.05&  \textbf{This work}\\     
&&&&&&&0.84$\pm$0.07&0.62$\pm$0.09&0.42$\pm$0.09&\textbf{This work}\\

(517717) 2015 KZ120 &85.5&0.82&46.73& 10.0& &&0.78$\pm$0.05&0.49$\pm$0.03&0.61$\pm$0.06& \textbf{This work}\\
&&&&&&&0.93$\pm$0.05&0.43$\pm$0.05&0.54$\pm$0.07&\textbf{This work}\\

1999 LE31&151.8&0.47&8.14&12.5&&&0.74$\pm$0.07&0.51$\pm$0.05
&0.49$\pm$0.05&J05\\ 
        &&&&&&&0.75$\pm$0.02&0.45$\pm$0.02&0.54$\pm$0.03&J05\\

2000 HE46 &158.6&0.9&23.7&14.8&&&0.87$\pm$0.06&0.55$\pm$0.07&0.40$\pm$0.06&J05\\       

2002 RP120 &119.1&0.95&53.71&12.3&0.10&14.6&0.83$\pm$0.03&0.47$\pm$0.03&0.52$\pm$0.03&J05\\ 

2008 KV42 &103.4&0.49&41.68&8.8&&&0.82$\pm$0.09&0.47$\pm$0.06&0.42$\pm$0.06&H12\\    
                        &&&&&&&&0.47$\pm$0.06&0.42$\pm$0.06&S10\\

2010 BK118&143.9&0.98&376.1&10.2&0.06&52.0&0.76$\pm$0.03&0.57$\pm$0.03&0.51$\pm$0.03&J15\\

2010 OM101  &118.8&0.92&26.1&17.0&0.04&3.2&0.81$\pm$0.04&0.53$\pm$0.04&&J15\\

2010 OR1 &143.9&0.92&26.9&16.2&0.11&2.3&0.76$\pm$0.03&0.52$\pm$0.09&0.48$\pm$0.04&J15\\
        &&&&&&&0.81$\pm$0.04&0.51$\pm$0.09&0.43$\pm$0.02&J15\\

2010 WG9 &70.4&0.65&53.2 &8.3&0.07&112.7&&&0.52$\pm$0.02&R13\\     
    &&&&&&&0.73$\pm$0.05&0.37$\pm$0.04&&J15\\

2012 DR30&78.0&0.98&1445.3&7.1&&&0.65$\pm$0.04&0.56$\pm$0.03&0.42$\pm$0.04& K13\\
&&&&&&&0.81$\pm$0.05&0.49$\pm$0.05&0.49$\pm$0.07&\textbf{This work}\\

2012 YO6 &106.9&0.48&6.3&14.7&&&&&0.34$\pm$0.05&J15\\

2013 NS11&130.3&0.79&12.6&13.6&0.03&15.2&0.74$\pm$0.02&0.56$\pm$0.04&0.52$\pm$0.08&J15\\

2013 YG48 &61.3&0.75&8.18 &17.2&&&0.77$\pm$0.02&0.55$\pm$0.02&&J15\\ 

A/2019 U5 &113.5&1.002& -- &9.7&&&0.75$\pm$0.04&0.42$\pm$0.04&0.54$\pm$0.05& \textbf{This work}\\
  &&&&&&& 0.68 $\pm$0.05&0.42$\pm$0.05&0.51$\pm$0.05 & \textbf{This work}\\

2020 EP&76.4&0.76&10.50&15.9&&&0.83$\pm$0.09&0.51$\pm$0.06&0.42$\pm$0.10&\textbf{This work}\\

       \hline
    \end{tabular}
    
    \begin{flushleft}
        $^{1}$ Absolute magnitudes are taken from MPC.
        
        $^{2}$ Albedo and diameters are from \citet{Licandro2016, Fernandez2005, Nugent2016,Bauer2013}.
        
        $^{3}$ 
        H12: \citet{Hainaut2012}, J05: \citet{Jewitt05}, J15: \citet{Jewitt15},
        K13: \citet{Kiss2013}, P13: \citet{Pinilla-Alonso13}, 
        R13: \citet{Rabinowitz2013}, S10: \citet{Sheppard2010}, T16: \citet{Tegler2016}.
        
\end{flushleft}   

\end{table*}

\begin{figure}
        \centering
        \includegraphics[width=\columnwidth]{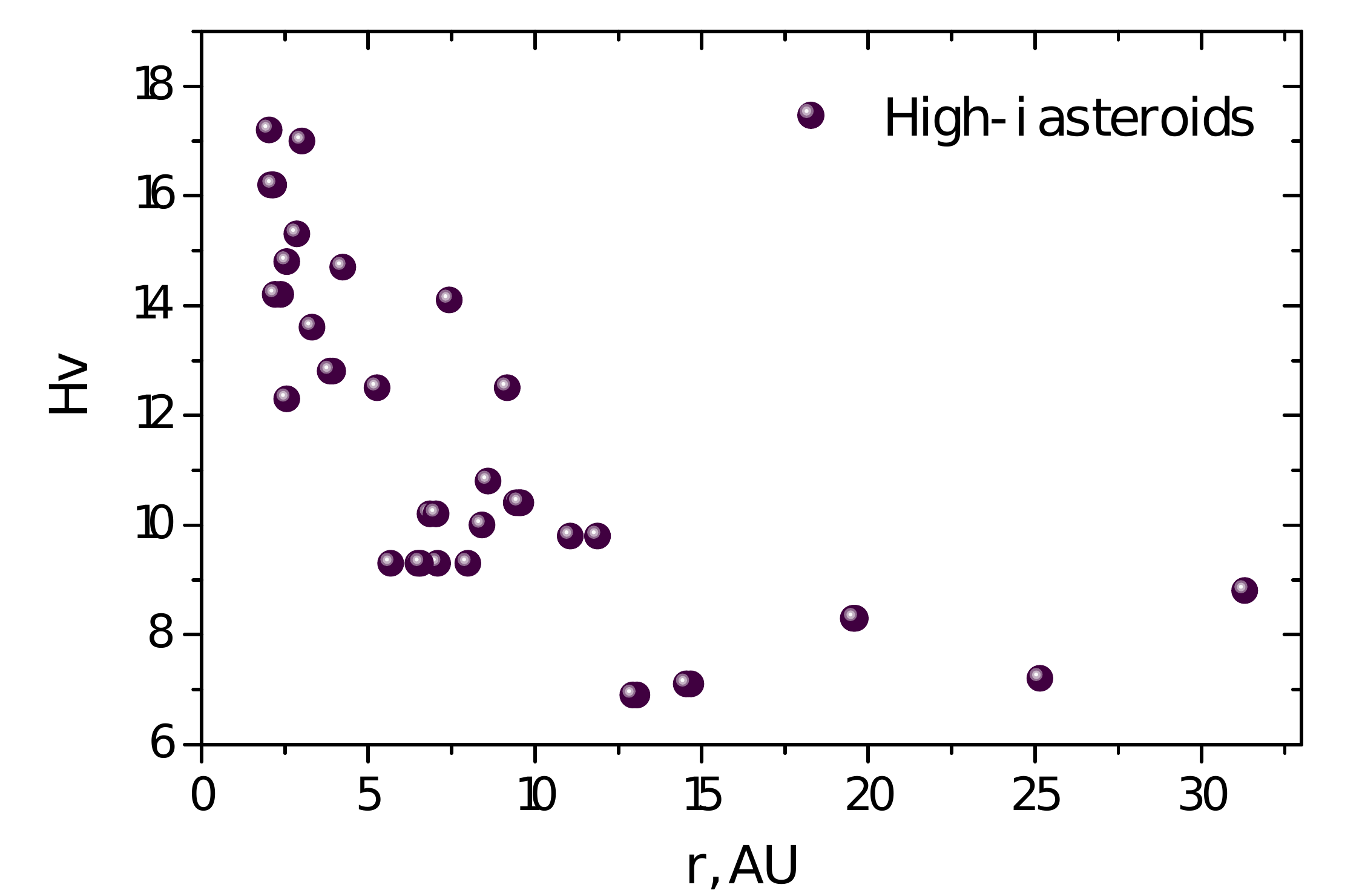}
        \includegraphics[width=\columnwidth]{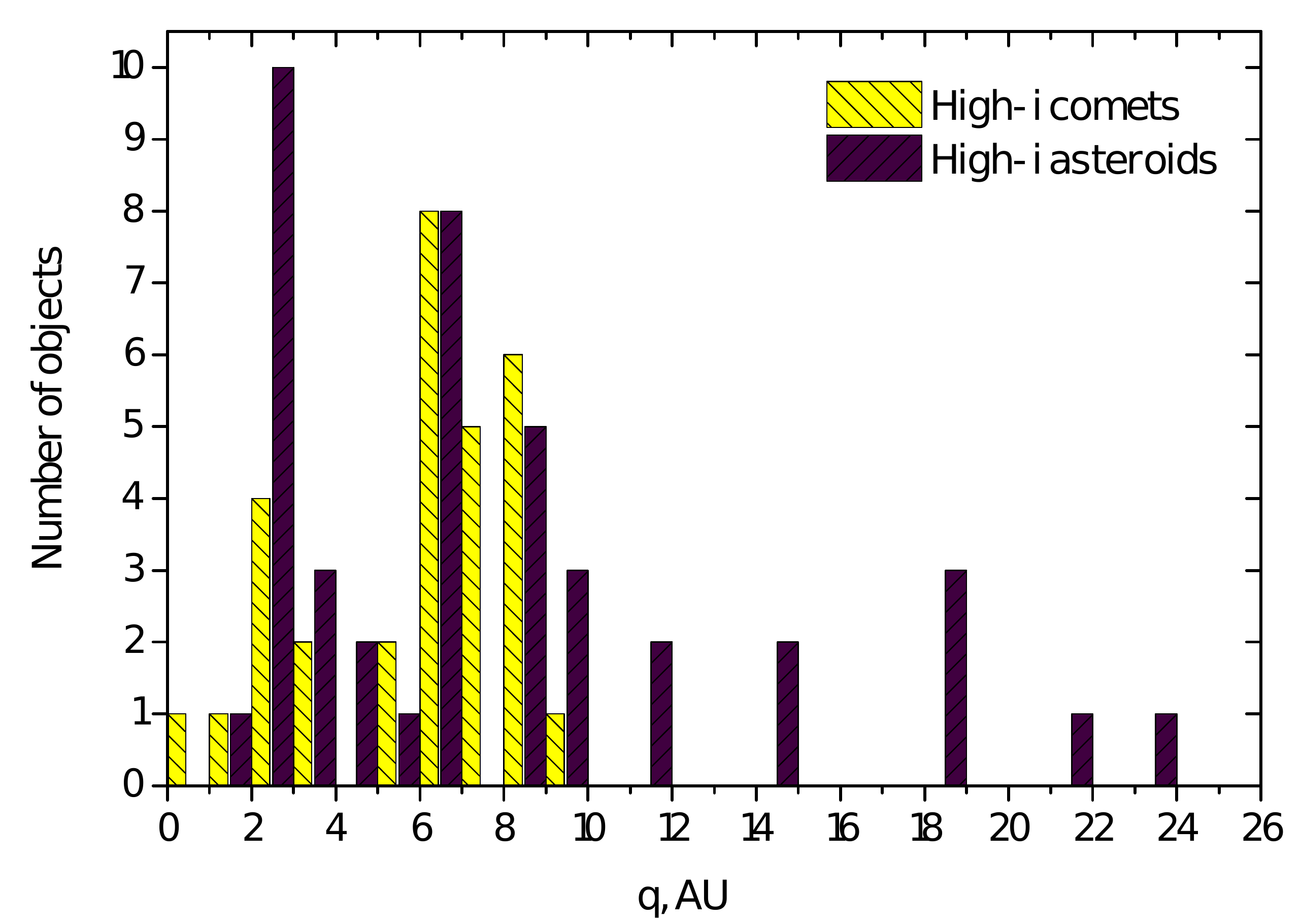}
        \caption{Upper panel: Absolute magnitude vs. heliocentric distance at the time of observation for high-inclination asteroids.
        Lower panel: Number of discovered high-inclination comets and asteroids observed at various perihelion distances.}
        \label{r_q_vs_Hv}
\end{figure}

As many of the high-inclination objects reside on cometary orbits, there is a question
of whether these objects are indeed comets with as-yet-undetected cometary activity.
We checked how close to the Sun these bodies were observed.
In Fig.~\ref{r_q_vs_Hv}~(upper panel) we show the absolute magnitude of high-inclination asteroids
and the heliocentric distances at which the corresponding objects were observed.
Most of the considered objects were observed only once, and the value of $r$ corresponds to the day of observation. 
When multiple observations were available, we chose the one that corresponds to the smaller value of $r$.
The lower panel of Fig. \ref{r_q_vs_Hv} shows a distribution of perihelion distances at the time of observations for both high-inclination asteroids and comets. 
From the histogram we can see that the two populations have similar distribution up to about 10~au (the only exception being a larger number of asteroids with $q$ around 3~au).
Thus, the figure shows that for the majority of the observed high-inclination asteroids, we cannot assume that the cometary activity was not detected due to the larger distance from the Sun.
If such asteroids are in fact dormant or dead comets, the absence of ultra-red matter can be associated with its blanketing, as we already discussed in this section.
However, there is a clear trend for smaller asteroids to be detected closer to the Sun, which may be caused by the observational bias.

\section{Lifetimes of high-inclination objects}

\subsection{Simulation of orbit lifetimes}

We simulated the orbital dynamics of the considered asteroids into the past and into the future, to estimate the time when they entered their retrograde orbits and when they will leave them. The simulations were conducted using the REBOUND package.\footnote{REBOUND code is freely available at http://github.com/hannorein/rebound} As a compromise between speed and accuracy, we used the WHFast integrator, included the perturbations from the four Jovian planets, and excluded the terrestrial planets. For each asteroid we created 96 clones with the normal distribution within the error ellipsoid of the orbital elements. The orbital elements and the covariation matrix of their errors were taken from JPL HORIZONS database.\footnote{JPL HORIZONS database is available at https://ssd.jpl.nasa.gov/horizons.cgi}

\begin{figure}
        \centering
        \includegraphics[width=\columnwidth]{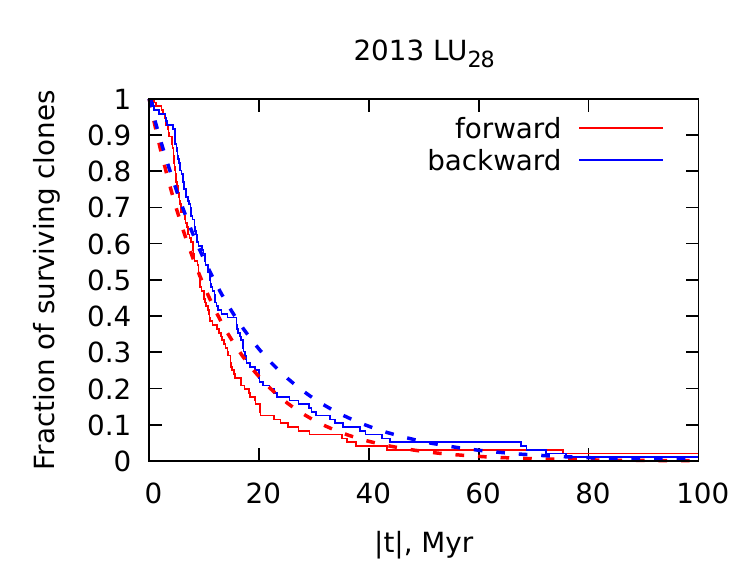}
        \includegraphics[width=\columnwidth]{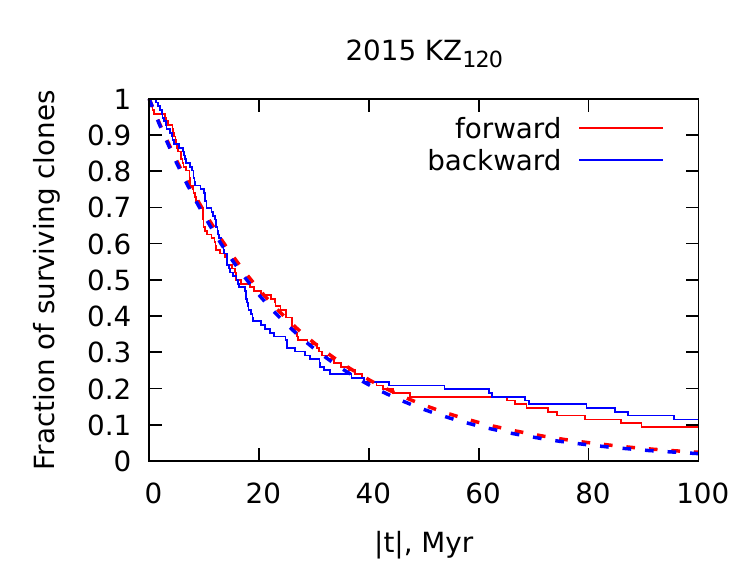}
        \includegraphics[width=\columnwidth]{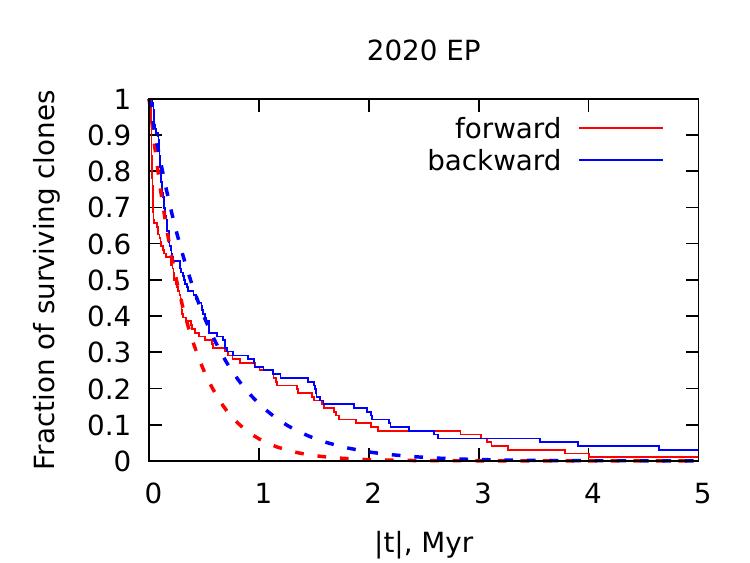}
        \caption{Fraction of surviving clones of asteroids 2013 LU28 (top),  2015 KZ120 (middle), and 2020 EP (bottom) as a function of time. The red and blue lines correspond to the forward and backward integration of orbits, respectively. The dashed lines show the exponential least-squares fits.}
        \label{fig-fraction}
\end{figure}

The results for the fraction of the surviving clones as a function of time are shown in Fig.~\ref{fig-fraction}. 
The figure shows  that there is no sharp moment when most clones leave the system: the population of surviving clones decays with time almost exponentially. 
This happens because the orbits of the considered objects are strongly chaotic, and the available precision of the orbital elements is insufficient to accurately predict their orbits, but can merely describe them statistically. 
In the vicinity of the nominal orbits there are some with very short and very long lifetimes, and each clone is randomly assigned a lifetime. 
In probabilistic terms, the nearly exponential decay implies that the probability for a clone to escape remains nearly constant with time.
Deviations from the exponential decay in Fig.~\ref{fig-fraction} include a small plateau for 2013 LU28 at small times (corresponding to relative stability of its present-day orbit) and a strong tails for 2015 KZ120 and 2020 EP at large times (corresponding to the existence of more stable orbits in  close vicinity to the nominal orbit).

The median lifetimes of the clones are given in Table \ref{tab-lifetime}. 
The median lifetime is easier to compute than the mean lifetime or its higher-order moments as it is less sensitive to the few longest surviving particles, and its computation does not require   waiting until all the particles leave the system. 
The statistical errors on the median lifetimes are computed, as  explained in Appendix A, based on the assumption of the exponential decay of the number of surviving particles with time.

We believe that the estimated future lifetimes of the clones are a good  statistical indicator of the expected future lifetime of the real asteroid, whereas the past lifetimes of the clones do not necessary indicate the real time passed by an asteroid on its orbit. The overwhelming majority of such simulated clones come from  interstellar space,  a highly improbable process due to the very low density of interstellar asteroids. Ignoring such clones and considering only the ones coming from the Kuiper belt or the main asteroid belt could significantly alter the lifetime distribution, but would be very hard to compute because of the extremely small number of such clones. A Bayesian analysis of the most plausible ways to enter the existing orbits is needed to estimate the past lifetimes more accurately.

\begin{table}
        \caption{Median lifetimes of asteroid clones simulated into the future and into the past. The computation of errors is explained in Appendix A.}
        \label{tab-lifetime}
        \begin{tabular}{l|c|c}
                & Lifetime & Lifetime \\
                Asteroid & in the past, Myr & in the future, Myr \\
                \hline
                (468861) 2013 LU28 & $-11.1 \pm 1.6$ & $9.2\pm 1.4$ \\ 
                (517717) 2015 KZ120 & $-16.1\pm 2.4$ & $16.3\pm 2.4$ \\
                2020 EP & $-0.32\pm 0.05$ & $0.23\pm 0.03$\\
        \end{tabular}
\end{table}

Lifetimes of 2013 LU28 can be compared with the calculations by \cite{Kankiewicz2017}. Assigning errors in accordance with our Appendix A, we can reword \cite{Kankiewicz2017} by saying that the lifetime in the past is $-8.9\pm 1.3$~Myr with the Yakovsky effect and $-9.2\pm 1.3$ for purely gravitational interactions; instead,  into the future the lifetimes are $8.4\pm 1.2$~Myr and $6.9\pm 1.0$~Myr with and without Yarkovsky, respectively. Although in all   four cases the times are somewhat shorter than our estimates,   the differences all fall within $1.5\sigma$, making them statistically insignificant.

\subsection{Influence of the Yarkovsky effect}

\begin{figure}
        \centering
        \includegraphics[width=\columnwidth]{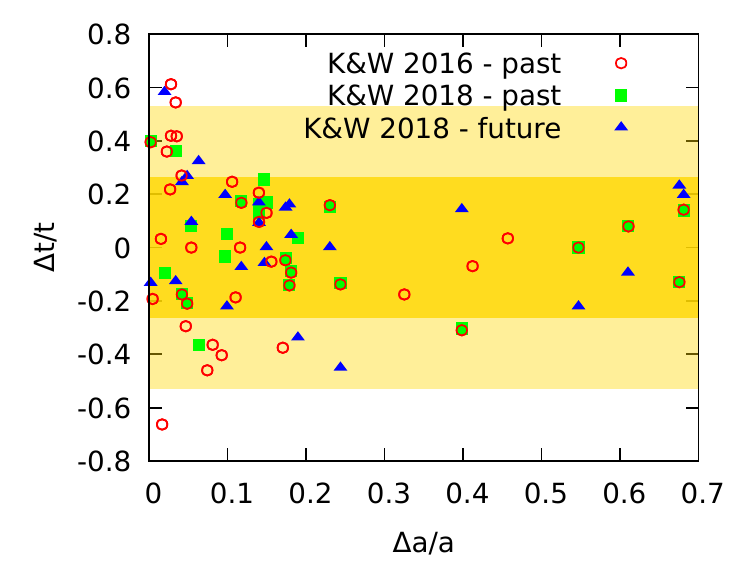}
        \caption{Relative change of the median escape times for asteroids due to the Yarkovsky effect. 
        Vertical coordinates are obtained by comparing computations with and without YORP for each asteroid from \cite{Kankiewicz2016} and \cite{Kankiewicz2018}. 
        Colour-coding distinguishes between  the two datasets, as well as integration into the past or   the future. 
        The horizontal axis is the relative distance to the closest planet from the asteroid's pericentre, apocentre, or semi-major axis (see text for details). The 1$\sigma$ and 2$\sigma$ areas for the shot noise are shaded in yellow.}
        \label{fig-points}
\end{figure}

The results with and without the Yarkovsky effect in \cite{Kankiewicz2017} seem  statistically indistinguishable in most cases. 
The relative deviations in the median lifetimes with and without the Yarkovsky effect are shown in Fig.~\ref{fig-points}.
Along the horizontal axis of this figure we plot $\Delta a/a$, where $a$ is the semi-major axis of an asteroid, and $\Delta a$ is determined as the minimum distance to a planet's orbit as measured from the asteroid's pericentre, apocentre, or semi-major axis (e.g. if the pericentre is 0.5~au   from Saturn, the apocentre 0.1~au from Neptune, and the semi-major axis 0.2~au from Uranus, then $\Delta a=0.1$~au).
The parameter  $\Delta a$ could serve as a crude indicator of the effect of a small change in  orbit  on the asteroid's dynamics. 
If a small perturbation of the asteroid’s orbit allows or forbids the intersection of  certain planetary orbits or puts the asteroid into a 1:1 resonance with a planet, then it can dramatically change the lifetime of the asteroid’s orbit. 
Otherwise small perturbations such as the Yarkovsky effect might be less significant.

We see in Fig.~\ref{fig-points} that for $\Delta a/a>0.05$ most points lie within the $1\sigma$ area, and all points lie within the $2\sigma$ area (in yellow). 
Only for the smallest $\Delta a/a$ do some points go beyond $2\sigma$.
Therefore, we conclude that for most asteroids the Yarkovsky effect changes the results on the level of the shot noise, thus it is indistinguishable from a random perturber. 
The few exceptions correspond to the orbits whose semi-major axis, aphelion, or perihelion are close to the orbit of one of the planets.

\subsection{Influence of the orbit uncertainty}

\begin{figure}
        \centering
        \includegraphics[width=\columnwidth]{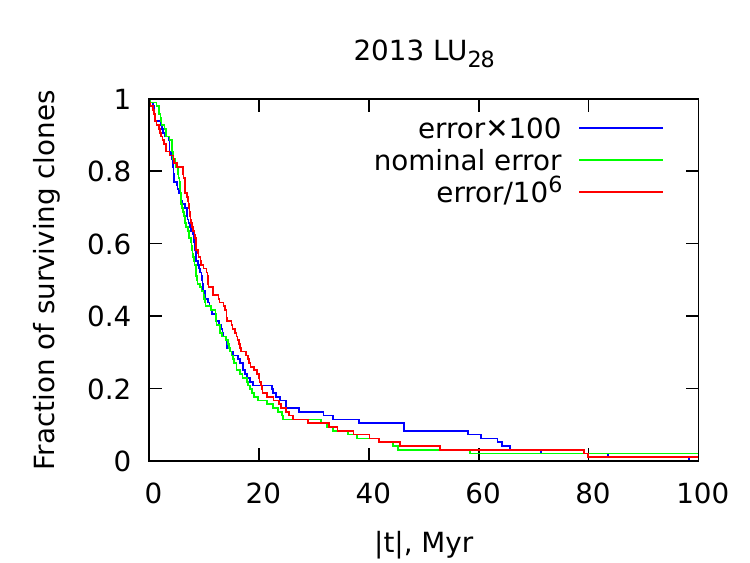}
        \includegraphics[width=\columnwidth]{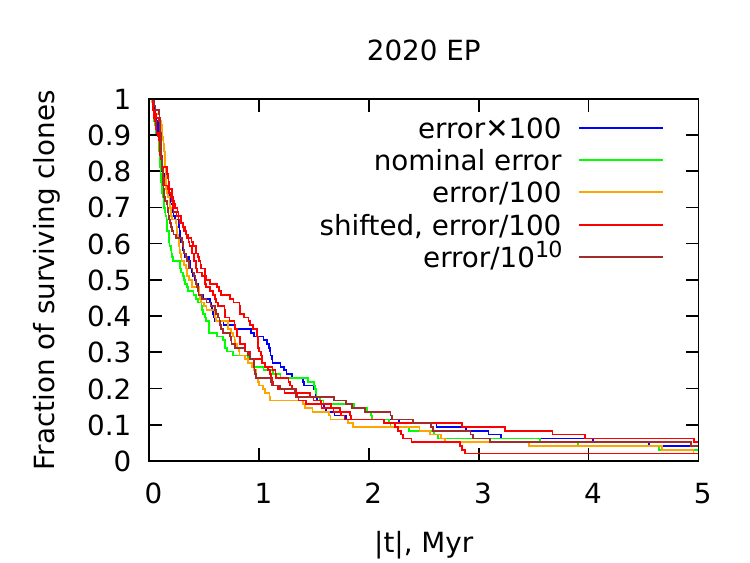}
        \caption{Distribution of asteroid clones over ejection times for the backward integration of asteroids (468861) 2013 LU28 (top) and 2020 EP (bottom) for artificially magnified, shrunk, or shifted error ellipsoids.}
        \label{fig-error}
\end{figure}

In Fig.~\ref{fig-error} we study how the distribution of asteroid clones over the lifetimes depends on the precision of the asteroid's orbit. 
We artificially multiply or divide errors on the orbital elements by some factor, and find that the lifetime distribution remains almost unaffected, even if the error is multiplied by 100 or divided by $10^{10}$. 
For 2020~EP we also try to shrink the error ellipsoid by the factor of 100, and then shift the median of the orbital element distribution to a random point within the old error ellipsoid (orange lines in the bottom panel), again with no significant change in the lifetime distribution.

In particular, this result implies that although the orbit of 2020 EP is very uncertain (the observation arc is only 54 days), the estimated lifetime will not change even if the accuracy of the orbital elements is improved by a significant factor.

This stability of the distributions seems less striking if we note that the Lyapunov times in the considered region of the solar system are typically much shorter than the orbit lifetimes. For example, for (468861) 2013 LU28 \citet{Kankiewicz2018} find the Lyapunov time to be  0.104 Myr, and the lifetime 6.9 Myr. 
The 66 times discrepancy between them translates into the error increment during the time of integration by the factor of $e^{66}\approx 10^{29}$. 
This factor rapidly turns a good knowledge of initial conditions into total unpredictability, and equates initial errors of $10^{-2}$ and $10^{-10}$. To have a chance of precisely predicting the orbit of 2013 LU28 throughout its entire lifetime, we   need the initial relative error in the orbital elements of the order of $10^{-29}$. 
Only then is there  a chance of obtaining in Figure~\ref{fig-error} a sharp drop at a certain instance of time instead of a dull exponent. 
Still, even with such an unrealistic accuracy of the initial conditions, for the precise prediction of the escape we will need an equally unrealistic accuracy, for example of the orbit integrator and planet masses. 
Even the Yarkovsky effect, although it gets lost in errors at the current accuracy of computations, will be crucial to know precisely if the other errors are eliminated.

 \section{Summary and conclusions}

We have presented new photometric observations for six high-inclination objects.
All of the observed targets have moderately red surfaces, which seem to be typical for high-inclination asteroids. 
Two of the targets in this work (C/2018~DO4 and C/2019 O3) appeared to be comets.
The level of dust production for comets was estimated by the means of parameter \textit{Af$\rho$}.
For C/2018~DO4 this value is around 100~cm at the aperture of about 6~000~km. 
The study of morphology for the comet C/2018~DO4  using different filtering techniques revealed some complex 
structures that may be associated with particles  that were ejected from the cometary nucleus.
The measured \textit{Af$\rho$} values for the comet C/2019 O3 are much larger than those for the comet C/2018~DO4 and are in a range from about 2~000 to 3~700~cm at   apertures from 30~000 to 60~000~km.

From the search in all-sky surveys, we were able to find some additional photometric data: four asteroids were found in the SDSS catalogue, and three in the Pan-STARRS images. 
However, due to the large error bars and scarcity of the found objects,
it did not help us to increase the statistics of high-inclination objects with knows surface properties.
Nonetheless, we can conclude that the mean colours of the found objects correspond to   moderately red surfaces. 

No confident detection of ultra-red objects were found in this work, which supports the results of \citet{Jewitt05, Jewitt15}.
The average colours of high-inclination objects are very similar to those of 
grey Centaurs and moderately red TNOs, and to a lesser extent to those of Jupiter trojans. 

For three high-inclination bodies, we simulated the orbital dynamics backward and forward in time,
and estimated the moment when these objects   entered and will leave their unstable orbits. 
The lifetimes in the forward and backward integrations both appeared strikingly insensitive to selection
of the orbital elements of the asteroid within their error margins, and to the value of the errors themselves. 
For example, increasing the error by a factor of 100 or decreasing it by a factor of $10^{10}$
produced virtually no change in the computed lifetime distribution of asteroid 
(468861) 2013~LU28. 
Evaluation of the shot noise of the median lifetime estimate allowed us to re-assess the previous claims 
\citep{Kankiewicz2017,Kankiewicz2018} about the influence of the Yarkovsky effect on the asteroid lifetime,
and in most cases we found its influence to be below the threshold of statistical significance.
It could imply that in such a strongly chaotic region of the solar system it is virtually impossible to conduct any reliable predictions of orbits into the distant past or future, but only very crude statistical estimates for ensembles of drastically different closely positioned orbits. 
These statistical estimates might be very robust with respect to small perturbations (such as the errors on the orbital elements or the Yarkovsky effect), but their predictive value is very limited.

In conclusion, our knowledge about high-inclination objects remains quite limited.
Thus, we are planning to continue the programme dedicated to the investigations of these rare solar system objects.

\begin{acknowledgements}

OI thanks the Slovak Academy of Sciences grant VEGA 2/0023/18 and the Slovak Research and Development Agency under the Contract no. APVV-19-0072. \\

Funding for the Sloan Digital Sky Survey IV has been provided by the Alfred P. Sloan Foundation, the U.S. Department of Energy Office of Science, and the Participating Institutions. SDSS acknowledges support and resources from the Center for High-Performance Computing at the University of Utah. The SDSS web site is www.sdss.org.
SDSS is managed by the Astrophysical Research Consortium for the Participating Institutions of the SDSS Collaboration including the Brazilian Participation Group, the Carnegie Institution for Science, Carnegie Mellon University, the Chilean Participation Group, the French Participation Group, Harvard-Smithsonian Center for Astrophysics, Instituto de Astrofísica de Canarias, The Johns Hopkins University, Kavli Institute for the Physics and Mathematics of the Universe (IPMU) / University of Tokyo, Lawrence Berkeley National Laboratory, Leibniz Institut für Astrophysik Potsdam (AIP), Max-Planck-Institut für Astronomie (MPIA Heidelberg), Max-Planck-Institut für Astrophysik (MPA Garching), Max-Planck-Institut für Extraterrestrische Physik (MPE), National Astronomical Observatories of China, New Mexico State University, New York University, University of Notre Dame, Observatório Nacional / MCTI, The Ohio State University, Pennsylvania State University, Shanghai Astronomical Observatory, United Kingdom Participation Group, Universidad Nacional Autónoma de México, University of Arizona, University of Colorado Boulder, University of Oxford, University of Portsmouth, University of Utah, University of Virginia, University of Washington, University of Wisconsin, Vanderbilt University, and Yale University.\\

The Pan-STARRS1 Surveys (PS1) and the PS1 public science archive have been made possible through contributions by the Institute for Astronomy, the University of Hawaii, the Pan-STARRS Project Office, the Max-Planck Society and its participating institutes, the Max Planck Institute for Astronomy, Heidelberg and the Max Planck Institute for Extraterrestrial Physics, Garching, The Johns Hopkins University, Durham University, the University of Edinburgh, the Queen's University Belfast, the Harvard-Smithsonian Center for Astrophysics, the Las Cumbres Observatory Global Telescope Network Incorporated, the National Central University of Taiwan, the Space Telescope Science Institute, the National Aeronautics and Space Administration under Grant No. NNX08AR22G issued through the Planetary Science Division of the NASA Science Mission Directorate, the National Science Foundation Grant No. AST-1238877, the University of Maryland, Eotvos Lorand University (ELTE), the Los Alamos National Laboratory, and the Gordon and Betty Moore Foundation.
\end{acknowledgements}

\bibliographystyle{aa} %
\bibliography{retro} %

\begin{appendix}
\section{Estimate of the error in the median orbit lifetime}

Here we show how we estimate the error on the median orbit lifetime. 
This estimate is useful not merely to establish confidence limits and compare our results to the predecessors, but also allows us to make more general conclusions regarding, for example,  the statistical insignificance of the lifetime alteration by the Yarkovsky effect for most retrograde asteroids.

\begin{figure}
\centering
\begin{tikzpicture}
\fill[color=blue!10] (2.08,0) rectangle (2.8,2.5);
\draw[-{Latex[length=3mm]}] (0,0)--(0,5.7);
\draw[-{Latex[length=3mm]}] (0,0)--(8,0);
\draw[samples=100,domain=0:2.5,color=red,thick] plot(3*\x,{5*exp(-\x)});
\draw[color=red,thick] (2.08,0)--(2.08,2.7);
\draw[color=blue,thick] (2.8,0)--(2.8,2.7);
\draw[-{Latex[length=3mm]},color=blue] (2.08,2.5)--(2.8,2.5);

\filldraw[color=blue,fill=blue!30,thick](1.602214916,2.1962944821) circle(0.08);
\filldraw[color=blue,fill=blue!30,thick](3.4210393093,1.5312519821) circle(0.08);
\filldraw[color=blue,fill=blue!30,thick](0.1232157683,3.0455722917) circle(0.08);
\filldraw[color=blue,fill=blue!30,thick](6.6846875007,0.2034983753) circle(0.08);
\filldraw[color=blue,fill=blue!30,thick](1.6863269982,1.0788898649) circle(0.08);
\filldraw[color=blue,fill=blue!30,thick](1.1903830402,1.889847579) circle(0.08);
\filldraw[color=blue,fill=blue!30,thick](0.923876026,1.2048496469) circle(0.08);
\filldraw[color=blue,fill=blue!30,thick](5.2963429535,0.0121378324) circle(0.08);
\filldraw[color=blue,fill=blue!30,thick](2.2027063258,0.157727078) circle(0.08);
\filldraw[color=blue,fill=blue!30,thick](1.2160814806,1.2998313166) circle(0.08);
\filldraw[color=blue,fill=blue!30,thick](0.9885275171,3.0893217804) circle(0.08);
\filldraw[color=blue,fill=blue!30,thick](0.4200143932,2.9346152565) circle(0.08);
\filldraw[color=blue,fill=blue!30,thick](6.0921327402,0.5036063958) circle(0.08);
\filldraw[color=blue,fill=blue!30,thick](2.2532223486,2.1487047431) circle(0.08);
\filldraw[color=blue,fill=blue!30,thick](5.4660412101,0.1743116731) circle(0.08);
\filldraw[color=blue,fill=blue!30,thick](0.4708825443,1.1560556416) circle(0.08);
\filldraw[color=blue,fill=blue!30,thick](1.0222993623,2.2811644398) circle(0.08);
\filldraw[color=blue,fill=blue!30,thick](2.7023969334,1.0279003685) circle(0.08);
\filldraw[color=blue,fill=blue!30,thick](5.5095803492,0.3348350029) circle(0.08);
\filldraw[color=blue,fill=blue!30,thick](0.8230233648,0.4227141727) circle(0.08);
\filldraw[color=blue,fill=blue!30,thick](2.6407803733,1.0662458239) circle(0.08);
\filldraw[color=blue,fill=blue!30,thick](1.7756379531,2.1554035938) circle(0.08);
\filldraw[color=blue,fill=blue!30,thick](4.3155783545,0.8028860209) circle(0.08);
\filldraw[color=blue,fill=blue!30,thick](3.0888931183,0.9427194293) circle(0.08);
\filldraw[color=blue,fill=blue!30,thick](3.3126148474,1.5770326676) circle(0.08);
\filldraw[color=blue,fill=blue!30,thick](2.6387082529,0.5314611355) circle(0.08);
\filldraw[color=blue,fill=blue!30,thick](0.1822095216,4.0746222657) circle(0.08);
\filldraw[color=blue,fill=blue!30,thick](2.6953420737,0.4147835643) circle(0.08);
\filldraw[color=blue,fill=blue!30,thick](0.1660931285,2.954174315) circle(0.08);
\filldraw[color=blue,fill=blue!30,thick](1.5668489034,1.9606904529) circle(0.08);
\filldraw[color=blue,fill=blue!30,thick](0.0991395589,0.3773930945) circle(0.08);
\filldraw[color=blue,fill=blue!30,thick](2.4250379986,0.8891328706) circle(0.08);
\filldraw[color=blue,fill=blue!30,thick](2.2780466906,1.1431206102) circle(0.08);
\filldraw[color=blue,fill=blue!30,thick](0.5002878321,3.7832165925) circle(0.08);
\filldraw[color=blue,fill=blue!30,thick](3.9849349681,0.6777042032) circle(0.08);
\filldraw[color=blue,fill=blue!30,thick](1.5876181256,1.5949324571) circle(0.08);
\filldraw[color=blue,fill=blue!30,thick](3.4576019989,1.4543871913) circle(0.08);
\filldraw[color=blue,fill=blue!30,thick](1.1661683996,2.3593227886) circle(0.08);
\filldraw[color=blue,fill=blue!30,thick](1.6513074419,2.6845788383) circle(0.08);
\filldraw[color=blue,fill=blue!30,thick](0.820396909,2.2127828926) circle(0.08);
\filldraw[color=blue,fill=blue!30,thick](5.6179899706,0.7180093068) circle(0.08);

\node[] at (0.6,5.5) {$f(t)$};
\node[] at (7.8,-0.3) {$t$};
\node[color=blue] at (2.44,2.8) {$\delta t$};
\node[color=blue] at (2.44,1.6) {$\delta N$};
\node[color=blue] at (2.9,-0.33) {$m_N$};
\node[color=red] at (2.08,-0.3) {$m$};

\end{tikzpicture}
\caption{Illustration of the sample median value. Theoretical distribution and its median value $m$ are in red. The sample points and the sample median value $m_N$ are in blue.}
\label{fig-median}
\end{figure}
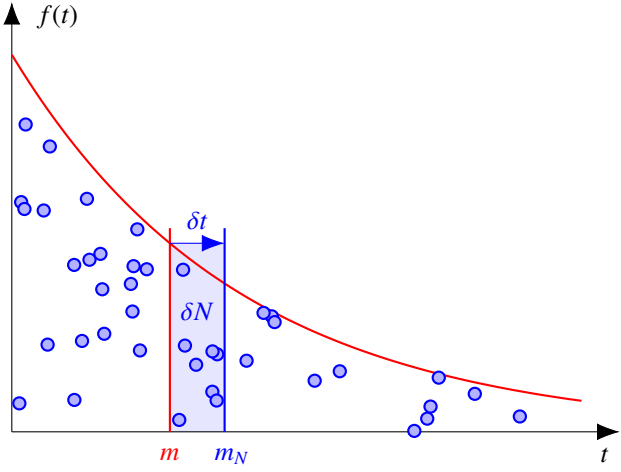

Consider a probability density function $f(t)=dP/dt$ of asteroid clones over lifetimes $t$. Let $m$ be the median value of this probability distribution, so that the probability of $t$ being smaller than $m$ is equal to one-half:
\begin{equation}
P(t<m)=\int_{0}^{m}f(t)dt=\frac{1}{2}.
\end{equation}
The probability density function and its median value are illustrated in Figure \ref{fig-median} with red lines.

Assume we are sampling the distribution over lifetimes for $N$ clones (small blue circles in Fig.~\ref{fig-median}). The number of clones below the theoretical median value $m$ is governed by the binomial distribution and may deviate from $\frac{N}{2}$ by some value $\delta N$. If $N$ is large, $\delta N$ is well approximated by the normal distribution, with the mean squared deviation 
\begin{equation}
\sigma_{\delta N}=\sqrt{\frac{1}{2}\cdot\frac{1}{2}\cdot N}=\frac{\sqrt{N}}{2}.
\label{sigmaN}
\end{equation}

When the sample median value $m_N$ is computed, it will be shifted with respect to $m$ by some value $\delta t=m_N-m$, such that the region between $m$ and $m_N$ accommodates $\delta N$ points. If $\delta t$ is small, the number of points between $m$ and $m_N$ can be approximated as the mathematical expectation of the number of points within the rectangle $\delta t\times f(m)$ (blue rectangle in Figure \ref{fig-median}):
\begin{equation}
\delta N\approx Nf(m)\delta t.
\label{dN-dx}
\end{equation}
Therefore, the distribution of $\delta t$ is proportional to the distribution of $\delta N$, and is thus also a normal distribution. Its mean squared deviation follows from Eq. (\ref{sigmaN}) and Eq. (\ref{dN-dx}):
\begin{equation}
\sigma_{\delta t}=\frac{1}{2\sqrt{N}f(m)}.
\label{sigma-general}
\end{equation}

For the strongly chaotic orbits of asteroid clones, we expect to have a more or less stable decay rate over time, with the probability density function approximately described by the decay exponent:
\begin{equation}
f(t)=\frac{1}{t_0}\exp\left(-\frac{t}{t_0}\right).
\label{f-exponent}
\end{equation}
This assumption is justified by Fig.~\ref{fig-fraction}, where  we see that the number of surviving clones is well approximated by an exponent. As the derivative of the number of surviving clones over time is proportional to the probability density function $f(t)$, the latter is thus also approximately exponential.

Then the median of such an exponential distribution is
\begin{equation}
m=t_0\log{2}.
\end{equation}
The error on the median as determined from Eq. (\ref{sigma-general}) is
\begin{equation}
\sigma_m=\frac{t_0}{\sqrt{N}}.
\end{equation}
Finally, the relative error is
\begin{equation}
\varepsilon_m=\frac{\sigma_m}{m}=\frac{1}{\sqrt{N}\log{2}}.
\end{equation}
This equation allows us to estimate the error on the median lifetime due to the shot noise, determined from the simulation of $N$ asteroid clones.

\end{appendix}

\end{document}